\documentclass[aps,prd,twocolumn,amsmath,superscriptaddress,amssymb,showpacs,floatfix,nofootinbib,longbibliography,preprintnumbers]{revtex4-1}
\pdfoutput=1
\usepackage{graphicx}
\usepackage{bm}
\usepackage{times}
\usepackage{slashed}
\usepackage{color}
\usepackage{slashed}
\usepackage{lipsum}
\usepackage{subfigure}
\usepackage{multirow}
\usepackage{amsmath}
\usepackage{array} 
\usepackage{varwidth} 

\usepackage{hyperref}
\hypersetup{
    pdfnewwindow=true,    
    colorlinks=true,      
    linkcolor=blue,       
    citecolor=blue,       
    filecolor=blue,      
    urlcolor=blue         
}

\newcommand{\be}{\begin{equation}}
\newcommand{\ee}{\end{equation}}
\newcommand{\bea}{\begin{eqnarray}}
\newcommand{\eea}{\end{eqnarray}}

\def\lsim{\mathrel{\raise.3ex\hbox{$<$\kern-.75em\lower1ex\hbox{$\sim$}}}}
\def\gsim{\mathrel{\raise.3ex\hbox{$>$\kern-.75em\lower1ex\hbox{$\sim$}}}}

\begin{document}

\preprint{FERMILAB-PUB-21-121-T}

\title{Evidence of TeV Halos Around Millisecond Pulsars}

\author{Dan Hooper}
\email{dhooper@fnal.gov,  ORCID: orcid.org/0000-0001-8837-4127}
\affiliation{Theoretical Astrophysics Group, Fermi National Accelerator Laboratory, Batavia, IL, 60510, USA}
\affiliation{Department of Astronomy \& Astrophysics and the Kavli Institute for Cosmological Physics (KICP), University of Chicago, Chicago, IL, 60637, USA}

\author{Tim Linden}
\email{linden@fysik.su.se, ORCID: orcid.org/0000-0001-9888-0971}
\affiliation{Stockholm University and The Oskar Klein Centre for Cosmoparticle Physics,  Alba Nova, 10691 Stockholm, Sweden}

\begin{abstract}

Using data from the HAWC gamma-ray Telescope, we have studied a sample of 37 millisecond pulsars (MSPs), selected for their spindown power and proximity. From among these MSP, we have identified four which favor the presence of very high-energy gamma-ray emission at a level of  $(2\Delta \ln \mathcal{L})^{1/2} \ge 2.5$. Adopting a correlation between the spindown power and gamma-ray luminosity of each pulsar, we performed a stacked likelihood analysis of these 37 MSPs, finding that the data supports the conclusion that these sources emit very high-energy gamma-rays at a level of $(2\Delta \ln \mathcal{L})^{1/2} = 4.24$. Among sets of randomly selected sky locations within HAWC's field-of-view, less than 1\% of such realizations yielded such high statistical significance. Our analysis suggests that MSPs produce very high-energy gamma-ray emission with a similar efficiency to that observed from the Geminga TeV-halo, $\eta_{\rm MSP} = (0.39-1.08) \times \eta_{\rm Geminga}$. This conclusion poses a significant challenge for pulsar interpretations of the Galactic Center gamma-ray excess, as it suggests that any population of MSPs potentially capable of producing the GeV excess would also produce TeV-scale emission in excess of that observed by HESS from this region. Future observations by CTA will be able to substantially clarify this situation.

\end{abstract}

\maketitle

\section{Introduction}

Observations by the High Altitude Water Cherenkov (HAWC) Observatory have identified bright, multi-TeV emission from the regions surrounding the nearby Geminga and Monogem pulsars~\cite{Albert:2020fua,Abeysekara:2017hyn,Abeysekara:2017old,Abdo:2009ku}. The spectra and intensity of these ``TeV halos'' indicate that roughly $\sim$\,$10\%$ of these pulsars' spindown power is converted into very high-energy electron-positron pairs. The angular extent of this emission (corresponding to roughly $\sim$\,$25 \,{\rm pc}$) indicates that cosmic-ray propagation is far less efficient in the vicinity of these pulsars than it is elsewhere in the interstellar medium~\cite{Hooper:2017gtd,Hooper:2017tkg,Johannesson:2019jlk,DiMauro:2019hwn,Liu:2019zyj}. 

In the time since HAWC's discovery of TeV halos around Geminga and Monogem, it has become increasingly clear that such emission is a nearly universal feature of middle-aged pulsars. In particular, a large fraction of the sources detected by HAWC~\cite{Albert:2020fua,Abeysekara:2017hyn,Smith:2020clm, Albert:2021vrd} (and many detected by HESS~\cite{H.E.S.S.:2018zkf,Abdalla:2017vci}) are spatially coincident with a pulsar. Moreover, there is a strong correlation between the spindown power of these pulsars and their observed gamma-ray luminosities. Recently, the HAWC Collaboration has produced a catalog of nine gamma-ray sources that have been detected at energies above 56 TeV~\cite{Abeysekara:2019gov}, all of which are located within $0.5^{\circ}$ of a known pulsar, most of which are very young and exhibit exceptionally high spindown fluxes (defined as the spindown power divided by the distance square). At present, all indications are that young and middle-aged pulsars are generically surrounded by spatially extended TeV halos, powered by the rotational kinetic energy of these objects, and which produce their observed gamma-ray emission through the inverse Compton scattering of very high-energy electrons and positrons on the surrounding radiation field~\cite{Linden:2017vvb,Sudoh:2019lav,Sudoh:2021avj}.

A more open question is whether recycled pulsars, with millisecond-scale periods, are also surrounded by TeV halos. Although no TeV sources are currently associated with a millisecond pulsar (MSP), this is not surprising given that even the brightest (highest spindown flux) MSPs would produce TeV halos that are only marginally detectable by existing TeV instruments  (assuming an efficiency similar to that of young and middle-aged pulsars). From a theoretical perspective, it is generally anticipated that MSPs (like young pulsars) should produce bright multi-TeV emission within their magnetospheres~\cite{Venter:2015gga,Bednarek:2016gpp,Venter:2015oza}, as their light curves indicate that the production of very high-energy electron-positron pairs is efficient~\cite{Venter:2015gga}. On the other hand, models of young and middle-aged pulsars generally include subsequent TeV-scale electron acceleration at the position of the pulsar wind nebula termination shock, a process which may only occur in the most powerful MSPs~\cite{2011ApJ...741...39S, Gaensler:2006ua}. Additionally, it is unclear whether or not diffusion is inhibited in the regions surrounding MSPs, as it is observed to be around young and middle-aged pulsars~\cite{Evoli:2018aza, Kun:2019sks}.

The question of whether MSPs generate TeV halos is important not only in terms of our understanding of the particle acceleration associated with these objects, but also with respect to the excess of GeV-scale gamma-ray emission that has been observed from the region surrounding the Galactic Center. The spectrum, morphology, and intensity of this excess agrees well with the predictions for annihilating dark matter particles~\cite{Hooper:2010mq, Daylan:2014rsa}. Alternatively, it has also been proposed that this excess emission could originate from a large population of unresolved MSPs, highly concentrated in the innermost volume of the Galaxy~\cite{Abazajian:2010zy, Lee:2015fea, Bartels:2015aea}. If it were confirmed that MSPs produce TeV halos (or other TeV-scale emission) at a level similar to young and middle-aged pulsars, measurements by HESS could be used to constrain the Inner Galaxy's MSP population to an abundance below that required to generate the GeV excess, potentially ruling out an MSP origin and providing significant support for the dark matter hypothesis.

In an earlier study~\cite{Hooper:2018fih}, we used an online tool released for public use by the HAWC Collaboration, based on the HAWC observatory's first 507 days of data (corresponding to the data set used to construct the 2HWC catalog~\cite{Abeysekara:2017hyn}), to search for evidence of very high-energy gamma-ray emission from MSPs. To this end, we performed an analysis of 24 MSPs within HAWC's field-of-view, identifying 2.6-3.2$\sigma$ evidence that these objects produce diffuse very high-energy gamma-ray emission consistent with TeV halo models. We found that these systems exhibit a similar efficiency (defined as the ratio of very high-energy gamma-ray emission to spindown power) to that required to explain the multi-TeV emission observed from Geminga, Monogem, and other middle-aged and young pulsars. 

In this paper, we revisit and expand upon this approach using an updated version of this online tool\footnote{\url{https://data.hawc-observatory.org/datasets/3hwc-survey/coordinate.php}} to study the HAWC data taken over its first 1523 days of operation (corresponding to the dataset used to construct the 3HWC catalog~\cite{Albert:2020fua}). Although the 3HWC catalog does not contain any sources that have been associated with an MSP~\cite{Albert:2020fua}, the approach pursued here allows us to look for TeV halos that do not necessarily meet the HAWC catalog's criteria of 5$\sigma$ statistical significance. Using the 3HWC Survey Tool, we make use of a larger collection of high-spindown power MSPs, utilize updated distance measurements, and take advantage of the larger HAWC data set. We find that this data supports the conclusion that MSPs emit very high-energy gamma-rays, and suggests that they produce this emission with a similar efficiency to that observed from the Geminga TeV-halo, $\eta_{\rm MSP} = (0.39-1.08) \times \eta_{\rm Geminga}$. This conclusion is difficult to reconcile with pulsar interpretations of the Galactic Center gamma-ray excess, as it indicates that any MSP population potentially responsible for the GeV excess would also produce TeV-scale emission at a level exceeding that observed from the region by HESS.

\section{Analysis Method}

In our analysis, we have included all MSPs found within the Australia Telescope National Facility (ATNF) pulsar catalog~\cite{Manchester:2004bp} that are located within HAWC's field-of-view ($-20^{\circ} < \delta < 50^{\circ}$), have a characteristic age greater than 1~Myr, and that exhibit a spindown flux greater than \mbox{$\dot{E}/d^2 > 5 \times~10^{33}$~erg/kpc$^2$/s.} The 37 MSPs that meet these requirements are shown in Table~\ref{TableList}, along with the values of their spindown power, distance, and spindown flux.

\begin{table}[t]
\begin{tabular}{|c|c|c|c|c|c|}
\hline 
Pulsar Name   & $\dot{E}$ (erg/s) & Dist.~(kpc) & $\dot{E}/d^2$~(erg/kpc$^2$/s) & $({\rm TS})^{1/2}$  
\tabularnewline
\hline
J0605+3757 & $9.3\times 10^{33}$ & 0.215 & $2.01 \times 10^{35}$ & -1.02 
\tabularnewline
\hline
J1400-1431 & $9.7\times 10^{33}$ & 0.278 & $1.26 \times 10^{35}$ & -1.04 
\tabularnewline
\hline
J1231-1411 & $1.8\times 10^{34}$ & 0.420 & $1.02 \times 10^{35}$ & -0.02 
\tabularnewline
\hline
J1737-0811 & $4.3\times 10^{33}$ & 0.206 & $1.01 \times 10^{35}$ & -0.16 
\tabularnewline
\hline
J1939+2134 & $1.1\times 10^{36}$ & 3.500 & $8.98 \times 10^{34}$ & 3.34 
\tabularnewline
\hline
J1710+4923 & $2.2\times 10^{34}$ & 0.506 & $8.59 \times 10^{34}$ & -0.62 
\tabularnewline
\hline
J1959+2048 & $1.6\times 10^{35}$ & 1.400 & $8.16 \times 10^{34}$ & 2.12 
\tabularnewline
\hline
J2214+3000 & $1.9\times 10^{34}$ & 0.600 & $5.28 \times 10^{34}$ & 0.33 
\tabularnewline
\hline
J1843-1113 & $6.0\times 10^{34}$ & 1.260 & $3.78 \times 10^{34}$ & 0.15 
\tabularnewline
\hline
J1300+1240 & $1.9\times 10^{34}$ & 0.709 & $3.78 \times 10^{34}$ & -0.59
\tabularnewline
\hline
J1744-1134 & $5.2\times 10^{33}$ & 0.395 & $3.33 \times 10^{34}$ & -0.95 
\tabularnewline
\hline
J0030+0451 & $3.5\times 10^{33}$ & 0.325 & $3.31 \times 10^{34}$ & 2.55 
\tabularnewline
\hline
J1023+0038 & $5.7\times 10^{34}$ & 1.368 & $3.05 \times 10^{34}$ & 2.56 
\tabularnewline
\hline
J2234+0944 & $1.7\times 10^{34}$ & 0.769 & $2.87 \times 10^{34}$ & 0.80 
\tabularnewline
\hline
J0218+4232 & $2.4\times 10^{35}$ & 3.150 & $2.42 \times 10^{34}$ & 1.56 
\tabularnewline
\hline
J0613-0200 & $1.3\times 10^{34}$ & 3.150 & $2.14 \times 10^{34}$ & 0.06 
\tabularnewline
\hline
J0337+1715 & $3.4\times 10^{34}$ & 1.300 & $2.01 \times 10^{34}$ & 0.25 
\tabularnewline
\hline
J1741+1351 & $2.3\times 10^{34}$ & 1.075 & $1.99 \times 10^{34}$ & 2.64 
\tabularnewline
\hline
J2339-0533 & $2.3\times 10^{34}$ & 1.100 & $1.90 \times 10^{34}$ & -0.35 
\tabularnewline
\hline
J0621+2514 & $4.9\times 10^{34}$ & 1.641 & $1.82 \times 10^{34}$ & 1.62
\tabularnewline
\hline
J0034-0534 & $3.0\times 10^{34}$ & 1.348 & $1.65 \times 10^{34}$ & 0.10 
\tabularnewline
\hline
J2042+0246 & $5.9\times 10^{34}$ & 0.640 & $1.44 \times 10^{34}$ & -0.67 
\tabularnewline
\hline
J1719-1438 & $1.6\times 10^{33}$ & 3.150 & $1.38 \times 10^{34}$ & -0.56 
\tabularnewline
\hline
J1921+1929 & $8.1\times 10^{34}$ & 2.434 & $1.37 \times 10^{34}$ & 0.62 
\tabularnewline
\hline
J1643-1224 & $7.4\times 10^{33}$ & 0.740 & $1.35 \times 10^{34}$ & 0.66 
\tabularnewline
\hline
J0023+0923 & $1.6\times 10^{34}$ & 1.111 & $1.30 \times 10^{34}$ & 1.33 
\tabularnewline
\hline
J2234+0611 & $1.0\times 10^{34}$ & 0.971 & $1.06 \times 10^{34}$ & -0.23 
\tabularnewline
\hline
J1911-1114 & $1.2\times 10^{34}$ & 3.150 & $1.05 \times 10^{34}$ & -0.02 
\tabularnewline
\hline
J1745-0952 & $5.0\times 10^{32}$ & 3.150 & $9.79 \times 10^{33}$ & -1.97 
\tabularnewline
\hline
J2256-1024 & $3.7\times 10^{34}$ & 2.083 & $8.53 \times 10^{33}$ & 0.45 
\tabularnewline
\hline
J1630+3734 & $1.2\times 10^{34}$ & 1.187 & $8.52 \times 10^{33}$ & -0.59 
\tabularnewline
\hline
J2017+0603 & $1.3\times 10^{34}$ & 1.399 & $6.64 \times 10^{33}$ & -0.37 
\tabularnewline
\hline
J1622-0315 & $8.1\times 10^{33}$ & 1.141 & $6.22 \times 10^{33}$ & -0.51 
\tabularnewline
\hline
J2043+1711 & $1.5\times 10^{34}$ & 1.562 & $6.15 \times 10^{33}$ & -0.72 
\tabularnewline
\hline
J0751+1807 & $7.3\times 10^{33}$ & 1.110 & $5.92 \times 10^{33}$ & -2.09 
\tabularnewline
\hline
J2302+4442 & $3.9\times 10^{33}$ & 0.859 & $5.29 \times 10^{33}$ & -0.01 
\tabularnewline
\hline 
J0557+1550 & $1.7\times 10^{34}$ & 1.834 & $5.05 \times 10^{33}$ & 0.14 
\tabularnewline
\hline
\end{tabular}
\caption{The 37 known millisecond pulsars in HAWC's field-of-view with a spindown flux ($\dot{E}/d^2$) greater than $5 \times 10^{33}$ erg/kpc$^2$/s. In addition to the spindown power and distance to each pulsar, in the rightmost column we show the statistical significance (the square root of the test statistic) in favor of very high-energy gamma-ray emission as detected by HAWC using their online 3HWC Survey Tool.}
\label{TableList}
\end{table}

For each of these MSPs, we have used the 3HWC Survey Tool to determine the test statistic (TS) for the hypothesis that there is a source of very high-energy gamma rays with a spectral slope of $-2.5$ in a given direction of the sky.\footnote{While the spectral index measured for Geminga's TeV halo is consistent with the value of $-2.5$ adopted in the 3HWC Survey Tool, Monogem's spectral index is somewhat harder, $\sim$\,$-2.0$.} The test statistic is defined in terms of the likelihood as ${\rm TS} \equiv 2 \ln \Delta \mathcal{L}$ and, in the absence of any gamma-ray sources, the TS will be distributed according to a $\chi^2$ distribution with one degree-of-freedom. The quantity $({\rm TS})^{1/2}$ then corresponds to the pre-trials significance in favor of a gamma-ray source being present in a given direction.

\begin{figure*}
\includegraphics[width=3.4in,angle=0]{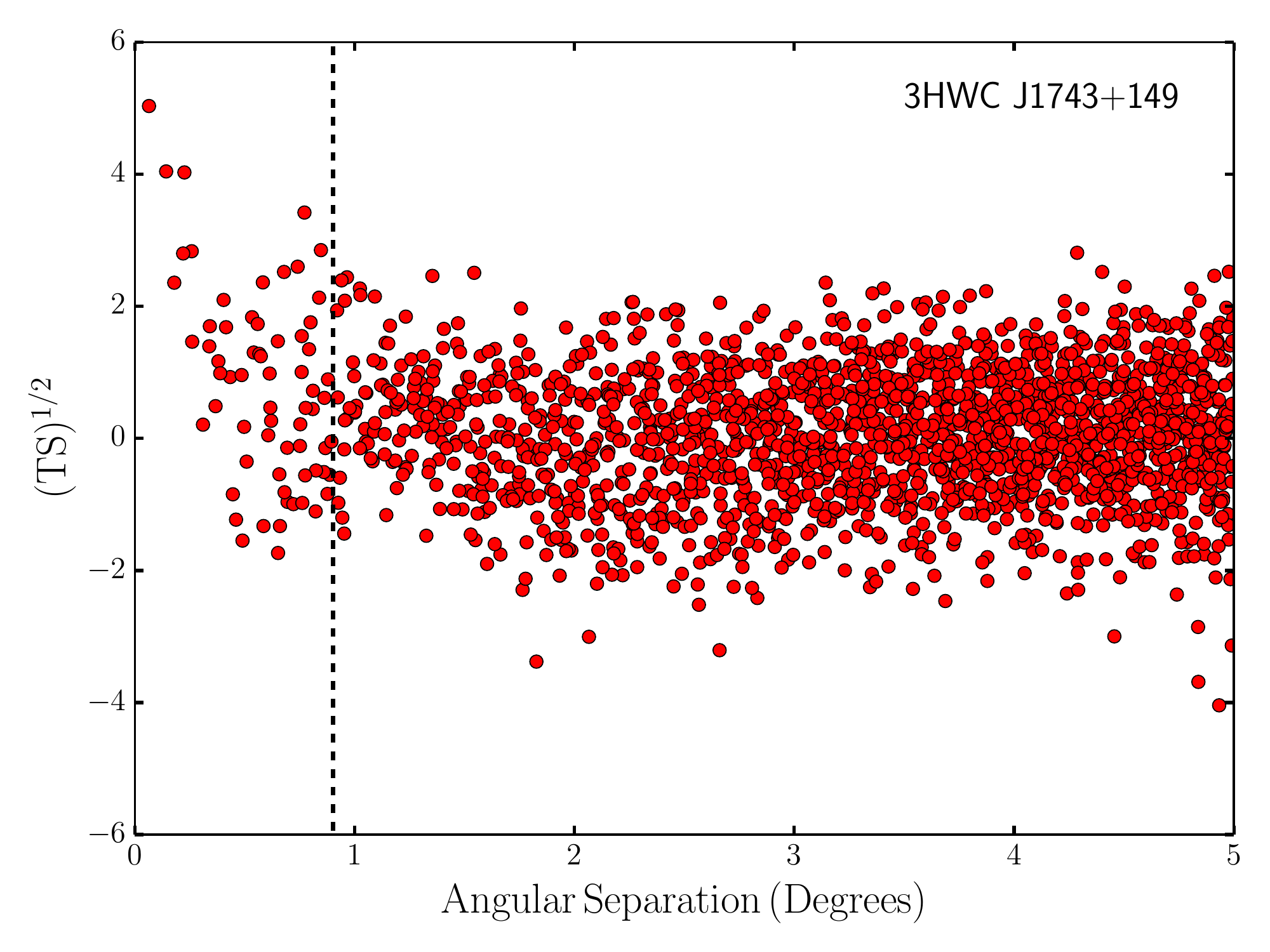}
\includegraphics[width=3.4in,angle=0]{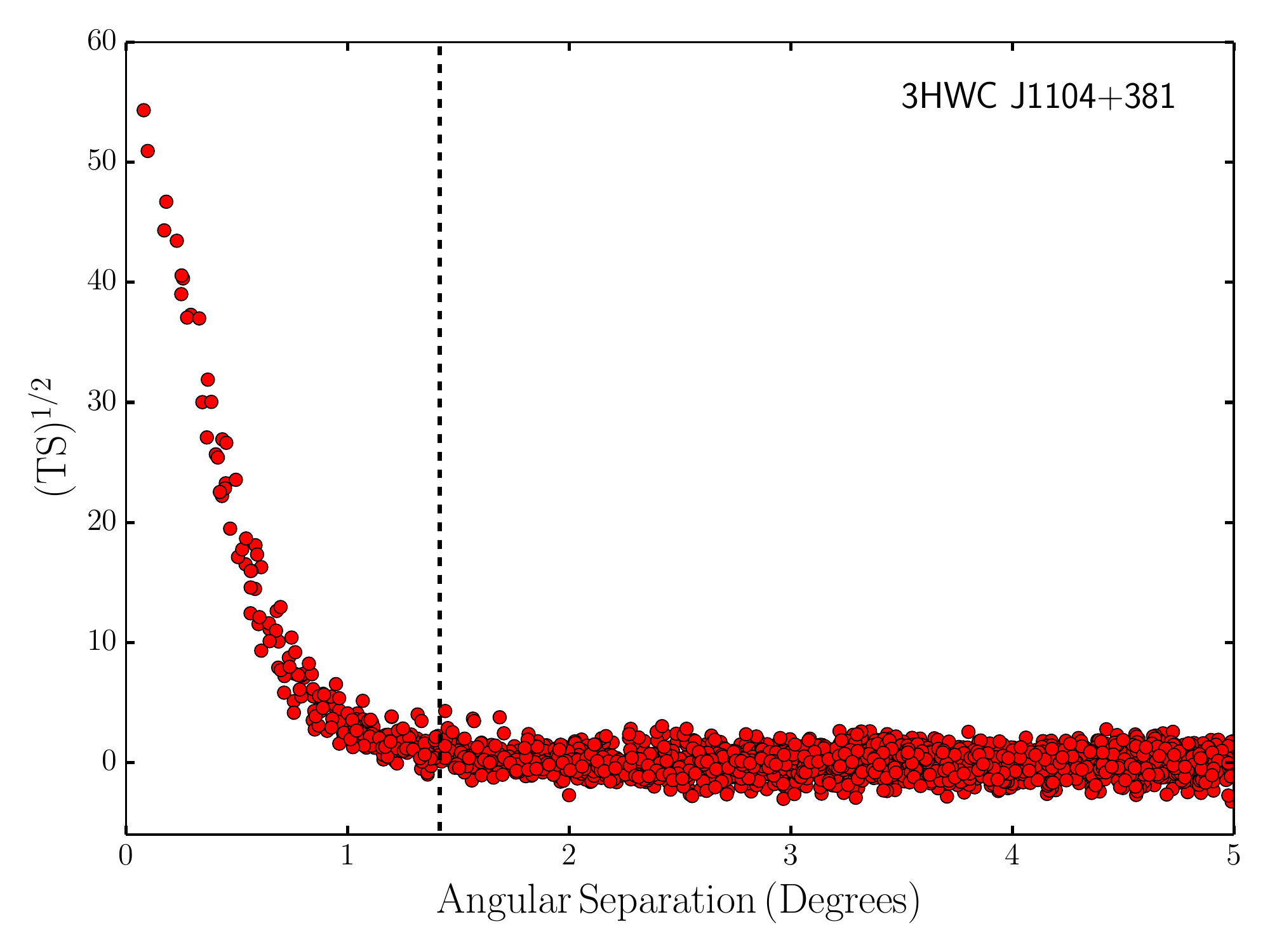}
\caption{The value of $(TS)^{1/2}$ found using the 3HWC Survey Tool in random sky locations near the faint HAWC source 3HWC J1743+149 (TS=25.9), and the bright HAWC source 3HWC J1104+381 (TS=3025.3). In removing any MSP from our analysis that is located within $0.5^{\circ} \times [\ln ({\rm TS}_{\rm 3HWC})]^{1/2}$ of any source in the 3HWC catalog (corresponding to the vertical dashed lines), we significantly limit the potential for contamination from this collection of sources.}
\label{nearbyTS}
\end{figure*}

The online HAWC Survey tool can be used to determine the TS value for four different spatial templates, which we select for each pulsar based on its distance. Operating under the assumption that typical TeV halos have a radius of $\sim$20-50~pc, we adopt the point-like template for pulsars with $d>2\,{\rm kpc}$, the $0.5^{\circ}$ extension template for $0.75 \, {\rm kpc} < d < 2 \, {\rm kpc}$, the $1^{\circ}$ template for $0.375 \, {\rm kpc} < d < 0.75 \, {\rm kpc}$, and the $2^{\circ}$ template for $d< 0.375 \, {\rm kpc}$. The square root of the test statistic for each of the 37 MSPs in our sample is reported in the rightmost column of Table~\ref{TableList}. Those entries which report a negative value for (TS)$^{1/2}$ represent directions of the sky which feature a statistical preference for a gamma-ray source with a negative normalization, likely resulting from an oversubtraction of the gamma-ray background.

In our previous study, in order to avoid regions of the sky that are contaminated by nearby gamma-ray sources, we considered only those MSPs that are more than $2^{\circ}$ from any point-like source, and more than $2^{\circ}$ from the edge of any extended source in the 2HWC catalog. Given the significantly larger number of sources in the 3HWC catalog (65, compared to 39 in the 2HWC catalog), such a cut would now remove several of the MSPs in our sample from consideration. With this in mind, we have adopted a cut designed to prevent any leakage greater than $\Delta {\rm TS} \gsim 1$ from any known source. Given that HAWC's point spread function is approximately $\sim 0.5^{\circ}$ over the energy range of interest, this cut requires us to remove from our sample any MSPs that are located less than $0.5^{\circ} \times [\ln ({\rm TS}_{\rm 3HWC})]^{1/2}$ from any source in the 3HWC catalog, where ${\rm TS}_{\rm 3HWC}$ is the test statistic of the source as reported by the HAWC collaboration. This cut removes from our analysis any MSP that is within $0.9^{\circ}$ of the faintest 3HWC sources (\textit{ie.} those with TS $\simeq$ 25), or within $1.6^{\circ}$ of the brightest 3HWC source (the Crab Nebula). The impact of this cut is illustrated in Fig.~\ref{nearbyTS}, for the representative HAWC sources 3HWC J1743+149 and 3HWC J1104+381. Although this cut does not remove from our analysis any of the 37 MSPs listed in Table~\ref{TableList}, it will impact later stages of our analysis.

From among the 37 MSPs considered in our analysis, we have identified four with $(TS)^{1/2} \ge 2.55$. Under the assumption that the background is Gaussian, the chance probability of identifying four such sources is 0.16\%, corresponding to a statistical significance 2.95$\sigma$. This is consistent with the results presented in our earlier analysis~\cite{Hooper:2018fih}. While this result is already interesting, it is potentially quite conservative, in that it weights all MSPs equally, regardless of their predicted gamma-ray flux (based on their spindown power and proximity).

\begin{figure}
\includegraphics[width=3.4in,angle=0]{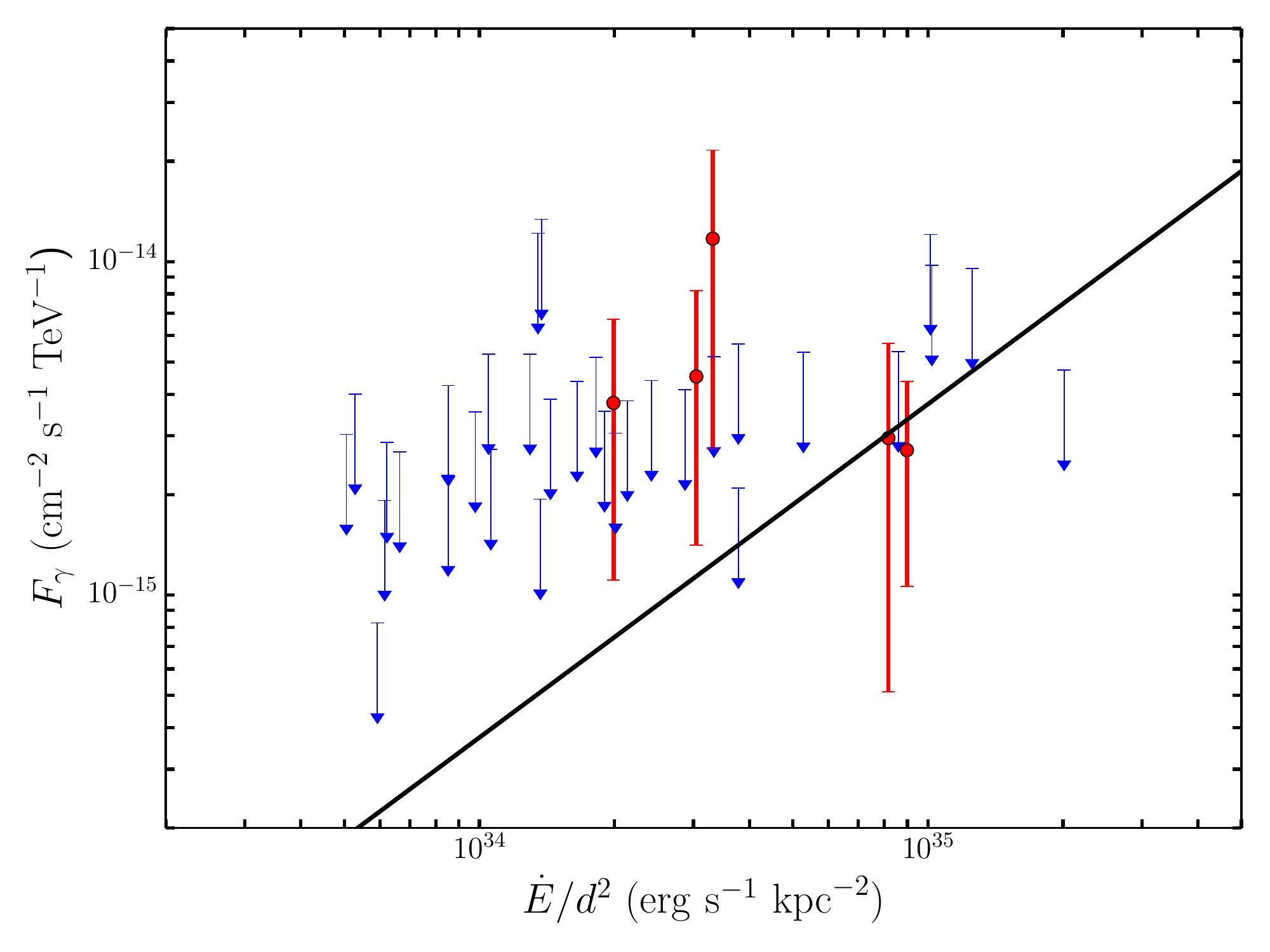}
\caption{The gamma-ray fluxes (evaluated at 7 TeV) of the 37 MSPs in our sample as determined using the 3HWC Survey Tool, and as a function of the spindown power, $\dot{E}/d^2$. For the 32 of these pulsars without any significant detection (those with $(TS)^{1/2}$ less than 2), we show the $2\sigma$ upper limit on the flux (blue). For each of the five sources with $(TS)^{1/2} > 2$, we plot the $1\sigma$ confidence interval on the gamma-ray flux (red). The black line represents the gamma-ray flux predicted under the assumption that each pulsar generates a TeV halo with the same efficiency as Geminga.}
\label{Edotflux}
\end{figure}

In Fig.~\ref{Edotflux}, we plot the gamma-ray fluxes (evaluated at 7 TeV) of the 37 MSPs in our sample as a function of $\dot{E}/d^2$, as determined using the 3HWC Survey Tool. For 32 of these pulsars (those with $(TS)^{1/2} < 2$), we show the $2\sigma$ upper limits on their flux. For each of the five sources with $(TS)^{1/2} > 2$, we plot the $1\sigma$ confidence interval. The black line in Fig.~\ref{Edotflux} represents the gamma-ray flux that is predicted under the assumption that each pulsar generates a TeV halo with the same efficiency as Geminga (\textit{ie.} that each pulsar has the same integrated gamma-ray flux above 0.1 TeV per unit spindown power as Geminga). In this regard, we have adopted here the following parameters for Geminga: $\dot{E}=3.25 \times 10^{34}$ erg/s, $d=0.19$ kpc~\cite{Manchester:2004bp}, and $dN_{\gamma}/dE_{\gamma} = 4.87 \times 10^{-14}$ TeV$^{-1}$ cm$^{-2}$ s$^{-1} \times (E_{\gamma}/7 \, {\rm TeV})^{-2.23}$~\cite{Abeysekara:2017hyn}. 

\begin{figure}
\includegraphics[width=3.4in,angle=0]{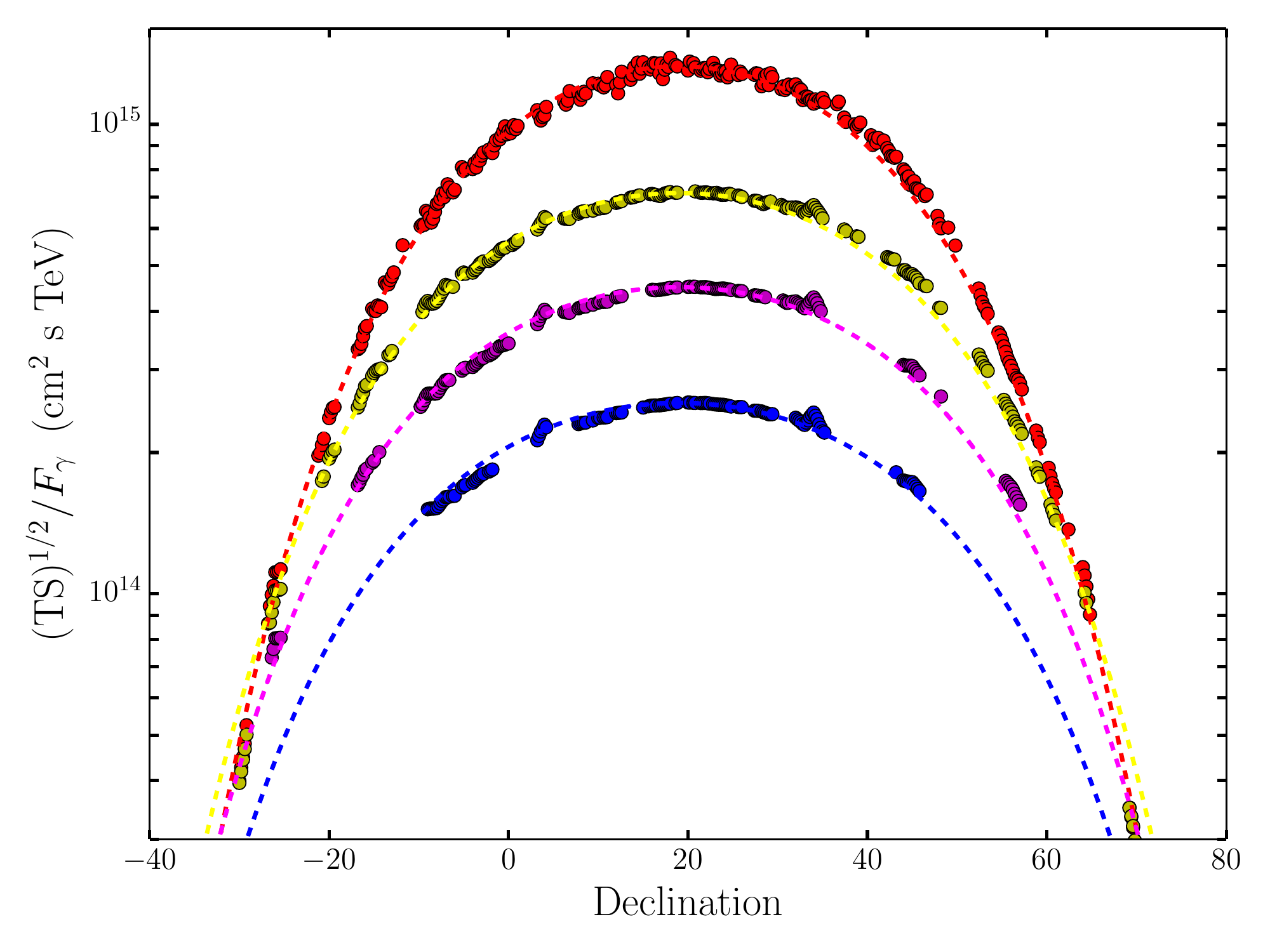}
\caption{The relationship between $({\rm TS})^{1/2}$/$F_{\gamma}$ and declination as found using the 3HWC Survey Tool on random sky locations (points). From top-to-bottom, the different colors correspond to the results obtained using the point source template, the $0.5^{\circ}$ extended template, the $1^{\circ}$ extended template, and the $2^{\circ}$ extended template. The dashed lines represent our polynomial fits to these results.}
\label{TSdec}
\end{figure}

The fact that 5 of the 18 highest spindown power MSPs within HAWC's field-of-view have yielded $2\sigma$ or higher evidence of gamma-ray emission is suggestive, but by no means overwhelming. If this population of sources has an approximately universal Geminga-like efficiency for gamma-ray emission, however, we should expect a joint analysis of these sources to provide a much more powerful test of this hypothesis. With this in mind, we have used the 3HWC Survey tool to calculate the likelihood that each of the 37 MSPs in our sample produces a very high-energy gamma-ray flux that is proportional to their spindown power, allowing the overall normalization of this proportionality to float.

We note that the 3HWC Survey Tool does not report negative values for fluxes, even when preferred by the data. Fortunately, there is reliable relationship between the quantity $({\rm TS})^{1/2}$/$F_{\gamma}$ and declination, which we use to determine the best-fit and limiting values of $F_{\gamma}$ in these instances (see Fig.~\ref{TSdec}). For each spatial extension template, we model this relationship as $\log_{10} [({\rm TS})^{1/2}/F_{\gamma}] = A +B \delta^2 +C \delta^4 +D \delta^6$, fitting each coefficient to the results of the 3HWC Survey Tool.

\begin{figure}
\includegraphics[width=3.4in,angle=0]{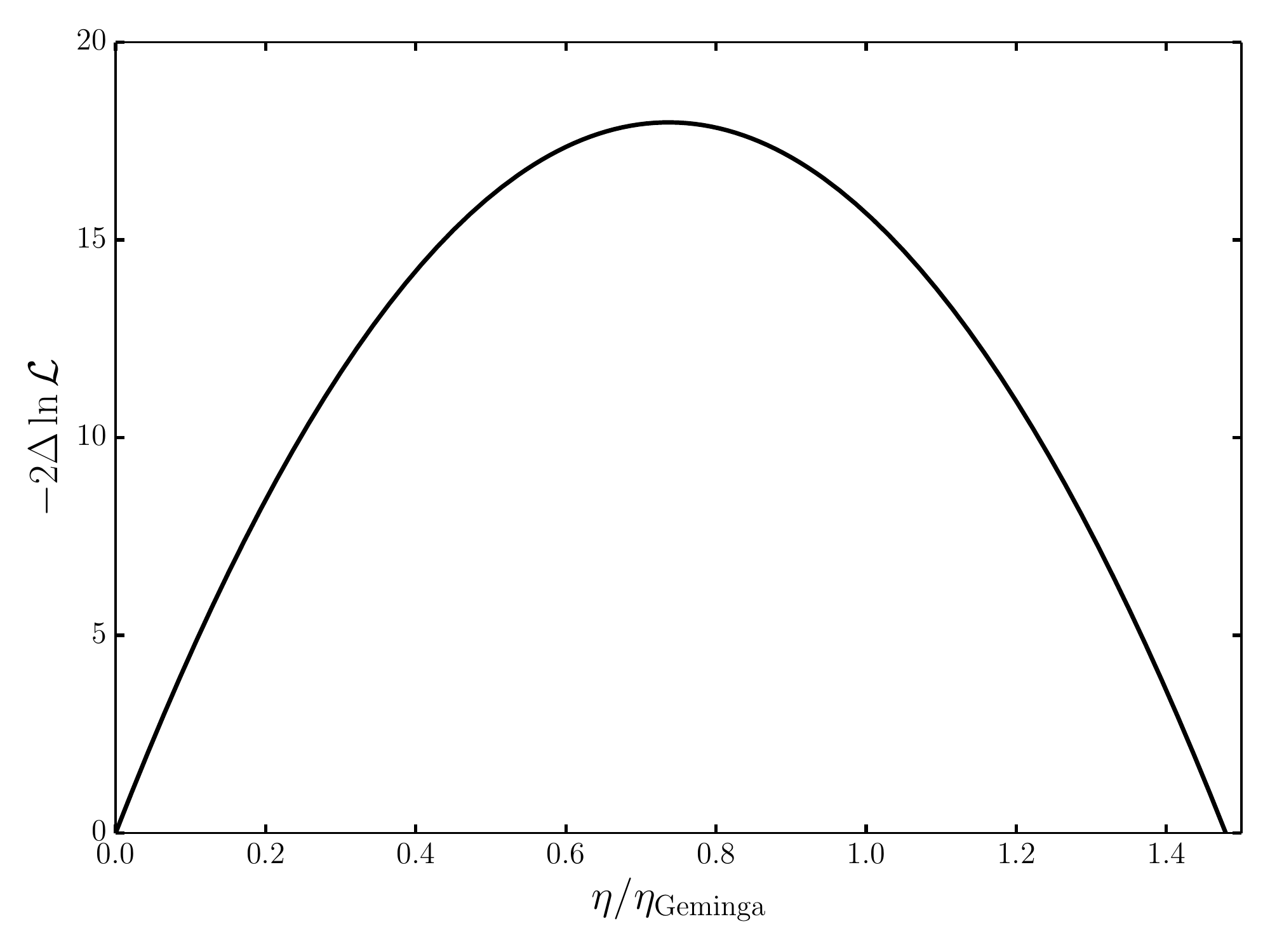}
\caption{The change to the log-likelihood (relative to no gamma-ray emission, $\eta_{\rm MSP}=0$) as a function of the efficiency for very high-energy gamma-ray production, assuming that all 37 MSPs have the same efficiency. The best-fit value of $\eta_{\rm MSP} = 0.74 \, \eta_{\rm Geminga}$ improves the fit by $2\Delta \ln \mathcal{L} = 17.97$, corresponding to a statistical preference of 4.24$\sigma$. At the $2\sigma$ level, our fit prefers $\eta_{\rm MSP} = (0.39-1.08) \times \eta_{\rm Geminga}$.}
\label{joint}
\end{figure}

The results of our joint likelihood is shown in Fig.~\ref{joint}. Under the assumption of a common gamma-ray efficiency for the 37 MSPs in our sample, we find that the HAWC data is best fit for an efficiency of $\eta_{\rm MSP} = 0.74 \, \eta_{\rm Geminga}$, which improves the fit by $2\Delta \ln \mathcal{L} = 17.97$ over the hypothesis of no gamma-ray emission from these sources, corresponding to a statistical preference of 4.24$\sigma$. At the $2\sigma$ level, our fit prefers an efficiency in the range of $\eta_{\rm MSP} = (0.39-1.08) \times \eta_{\rm Geminga}$.

\begin{figure}
\includegraphics[width=3.4in,angle=0]{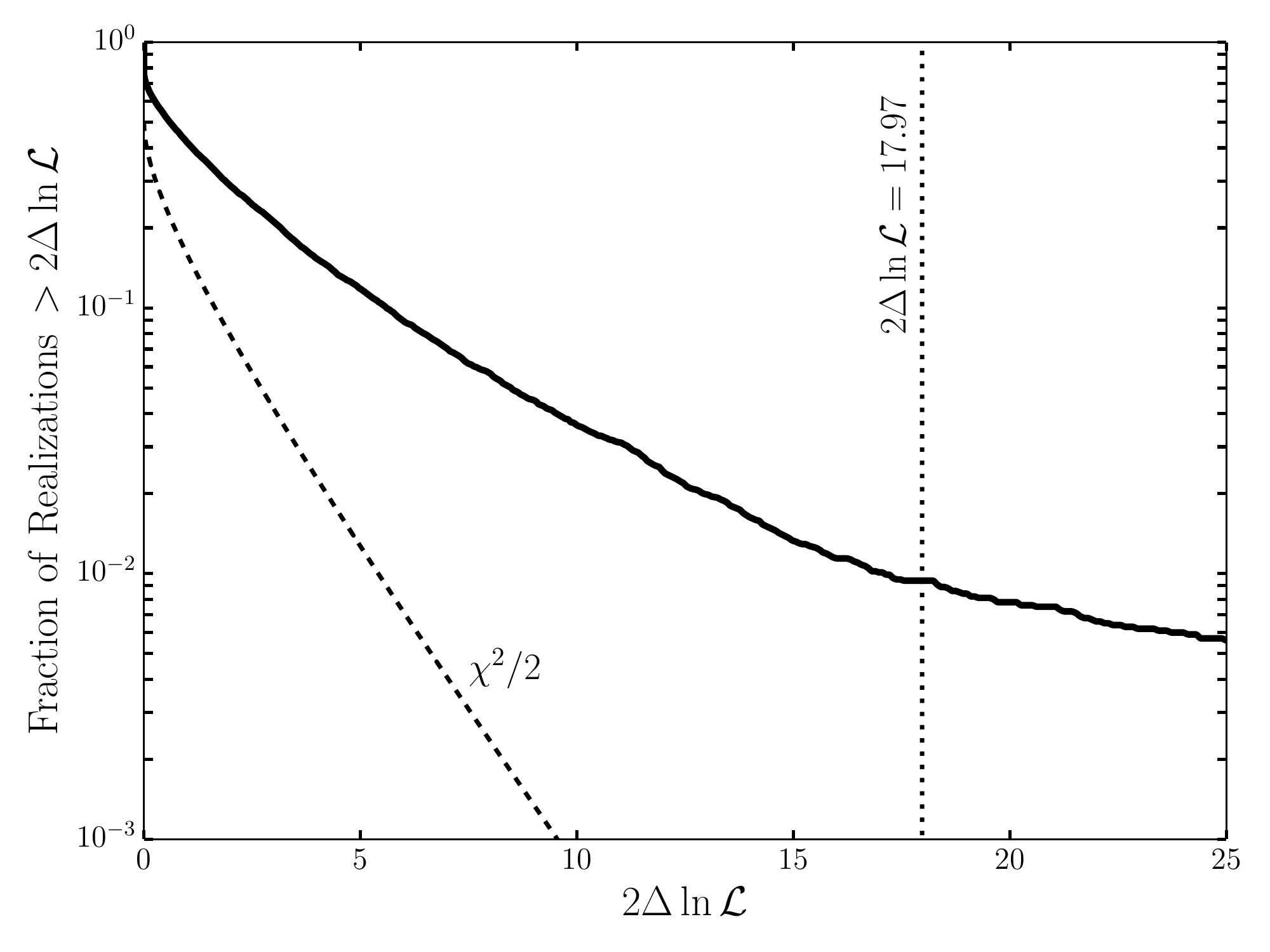}
\caption{The fraction of 10,000 randomly selected control group realizations that yield evidence for very high-energy gamma-ray emission at a given statistical significance (solid curve). This is compared to the predictions expected from a normal distribution, labeled $\chi^2/2$. From among this collection of 10,000 control group samples, only 94 (0.94\%) favored a non-zero gamma-ray flux with as much statistical significance ($2\Delta \ln \mathcal{L}=17.97$) as the collection of 37 MSPs considered in this study.}
\label{randomskies}
\end{figure}

As presented in the preceding paragraph, the statistical significance of our results depends on the assumption that the backgrounds are normally distributed. This is not, in fact, the case, and non-Gaussian tails are empirically observed in the distribution of TS values of random sky locations obtained using the 3HWC Survey Tool. With this in mind, we have constructed a control group by measuring the TS values of a large number of sky locations using the 3HWC Survey Tool. These ``blank sky" locations were selected such that they are each within HAWC's field-of-view and located no closer than $0.5^{\circ} \times [\ln ({\rm TS}_{\rm 3HWC})]^{1/2}$ from any source in the 3HWC catalog. To make this collection of sky locations most closely reflect the characteristics of our MSP sample, we have divided our 37 sources into $10^{\circ}$ bins in RA and Dec, and drawn control group sky locations from this statistical distribution of bins (and using a flat distribution within each $10^{\circ}$ increment). We have also randomly selected the distance to each control group source (this determines which angular template is used in the 3HWC Survey Tool), weighted by the distances to the 37 MSPs in our sample. From this distribution we generated 10,000 sets of 37 sky locations (and angular templates). From among this collection of 10,000 control group samples, only 94 (0.94\%) favored a non-zero gamma-ray flux at a level exceeding the statistical significance as our 37 MSPs (see Fig.~\ref{randomskies}).

\section{Implications for the Galactic Center Gamma-Ray Excess}

A bright and statistically significant excess of GeV-scale gamma rays from the region surrounding the Galactic Center has been identified in data collected by the Fermi Gamma-Ray Space Telescope~\cite{Daylan:2014rsa,TheFermi-LAT:2017vmf} (for earlier work, see Refs.~\cite{Goodenough:2009gk,Hooper:2010mq,Hooper:2011ti,Abazajian:2012pn,Hooper:2013rwa,Gordon:2013vta,Calore:2014xka,TheFermi-LAT:2015kwa}). The fact that the spectrum, morphology, and intensity of this excess are each consistent with the expectations from annihilating dark matter particles has generated a great deal of interest~\cite{Berlin:2014tja,Berlin:2014pya,Abdullah:2014lla,Hooper:2012cw,Martin:2014sxa,McDermott:2014rqa,Cahill-Rowley:2014ora,Hooper:2014fda,Liu:2014cma,Caron:2015wda,Berlin:2015wwa,Cline:2015qha,Bertone:2015tza,Fonseca:2015rwa,Freese:2015ysa,Alves:2014yha,Agrawal:2014una,Izaguirre:2014vva,Ipek:2014gua,Tang:2015coo,Escudero:2016kpw,Escudero:2017yia}. The primary alternative explanation for this signal is that these gamma rays are instead generated by a large population of centrally concentrated and unresolved MSPs~\cite{Hooper:2010mq,Abazajian:2010zy,Hooper:2011ti,Abazajian:2012pn,Hooper:2013rwa,Gordon:2013vta, Lee:2015fea, Bartels:2015aea}. 

To date, no individual MSPs have been detected in the Inner Galaxy. Despite this, interest in the possibility that MSPs could generate this signal grew considerably in 2015, when two groups claimed to find evidence of small-scale power in the excess, favoring point source interpretations of this signal~\cite{Lee:2015fea,Bartels:2015aea}. It was subsequently shown, however, that analyses making use of non-Poissonian templates (as were used in Ref.~\cite{Lee:2015fea}) tend to misattribute smooth gamma-ray signals to point source populations, especially in the presence of imperfectly modelled diffuse backgrounds~\cite{Leane:2019xiy, Leane:2020pfc,Leane:2020nmi}. Similarly, it was shown in Ref.~\cite{Zhong:2019ycb} that the small-scale power identified in the wavelet-based analyses of Ref.~\cite{Bartels:2015aea} is not attributable to the gamma-ray excess. At this time, the Fermi data cannot be said to favor a pulsar interpretation of this signal, as was previously claimed. Instead, this class of analyses can, at present, only be used to place constraints on the luminosity function of any point source population that might be present in the Inner Galaxy (see, for example, Refs.~\cite{List:2020mzd,Calore:2021bty}).

More recently, claims have been made that the gamma-ray excess is better fit by a template that reflects the spatial distribution of stars in the Milky Way's bulge and bar than that of a spherically symmetric, dark matter-like template~\cite{Macias:2016nev,Bartels:2017vsx,Macias:2019omb}. If confirmed, this would favor scenarios in which the gamma-ray excess is generated by MSPs or other point sources which trace the stellar distribution of the Inner Galaxy. The conclusions of these studies, however, are highly sensitive to the details of the background model adopted, and on the spatial tails of the excess. Given the uncertainties associated with these factors, it is far from clear which of these templates better resembles the morphology of the actual gamma-ray excess. 

Several arguments have been leveled against MSP interpretations of the gamma-ray excess. Firstly, if the central MSP population exhibited the same gamma-ray luminosity function as those observed in the disk~\cite{Cholis:2014lta,Hooper:2015jlu,Bartels:2017xba} and globular cluster populations~\cite{Hooper:2016rap} of the Milky Way, Fermi should have already resolved $\sim \mathcal{O}(10-20)$ individual MSPs from the Inner region of the Galaxy. In contrast, no such sources have been identified. Furthermore, the number of low-mass X-ray binaries observed in the Inner Galaxy suggests that no more than $\sim$\,$4-23\%$ of the gamma-ray excess originates from MSPs~\cite{Haggard:2017lyq} (see also, Ref.~\cite{Cholis:2014lta}).

Looking forward, there are a number of ways in which this situation could be substantially clarified. First, if MSPs are responsible for this excess, a significant number of these objects should be detectable with upcoming large-area radio surveys~\cite{Calore:2015bsx}. Second, if gamma-ray emission with the same spectral shape as observed from the Galactic Center is detected from one or more dwarf galaxies, this would provide a strong confirmation of the dark matter interpretation of this signal~\cite{Fermi-LAT:2016uux}. This appears particularly promising in light of the large number of dwarf galaxies that the Rubin Observatory is anticipated to discover. Third, if AMS-02 were to robustly confirm the presence of the cosmic-ray antiproton excess reported in Refs.~\cite{Cuoco:2016eej,Cui:2016ppb,Cholis:2019ejx,Cuoco:2019kuu}, this would also serve to bolster the dark matter interpretation of the Galactic Center excess.

The results of this study have significant bearing on the question of the origin of the Galactic Center gamma-ray excess. In particular, if MSPs generate TeV halos with a Geminga-like efficiency, as our results indicate is the case, then any MSPs which contribute to the Fermi excess should also produce significant emission at very-high energies. In our previous study~\cite{Hooper:2018fih}, we showed that if there are enough MSPs in the Inner Galaxy to produce the Fermi excess, their TeV emission (if produced with a Geminga-like efficiency) would exceed or saturate that observed by HESS from the innermost $0.5^{\circ}$ around the Galactic Center~\cite{Abramowski:2016mir}. While this TeV emission could plausibly be suppressed by the presence of strong magnetic fields, this would result in more radio synchrotron emission than is observed~\cite{Hooper:2018fih}. While we do not repeat here the calculations which support this conclusion, the results shown in Figs.~4 and~5 of Ref.~\cite{Hooper:2018fih} (see also, Ref.~\cite{Hooper:2017rzt}) apply if MSPs have Geminga-like TeV halos, as is indicated by results presented in this study.\footnote{Similar conclusions were reached in Ref.~\cite{Song:2019nrx}, in particular in the case that TeV halos inject electrons and positrons with a spectral index similar to that observed from Geminga and Monogem, $\Gamma \sim -1.5$ to $-2$.}

Lastly, we note that while measurements from HESS currently place strong constraints on the TeV flux (and thus the total number of TeV-emitting MSPs) within the inner few degrees around the Galactic Center, this region is somewhat smaller than the $\sim$5-10$^\circ$ region in which the GeV excess is most pronounced. Fortunately, upcoming observations by the Cherenkov Telescope Array (CTA)~\cite{CTAConsortium:2018tzg} will be able to produce high-resolution maps of the very high-energy gamma-ray emission from the entire Inner Galaxy. By either identifying the TeV halo emission from a large MSP population, or by placing strong constraints on the number of MSPs present in the Inner Galaxy, CTA is expected to be able to clarify this situation considerably~\cite{Macias:2021boz}.

\section{Summary and Conclusions}

In this study, we have presented significant evidence that millisecond pulsars (MSPs), like young and middle-aged pulsars, produce and are surrounded by TeV halos. Using data provided by the HAWC Collaboration, we performed a stacked likelihood analysis of 37 MSPs selected for their spindown power and proximity, finding evidence that these objects produce very high-energy gamma-ray emission, with a statistical significance corresponding to $(2\Delta \ln \mathcal{L})^{1/2} = 4.24$. Using sets of randomly selected sky locations as a control group, we found that less than 1\% of such realizations yielded as much statistical significance. 

Our analysis indicates that MSPs produce very high-energy gamma-ray emission with a similar efficiency to that observed from the Geminga TeV-halo, $\eta_{\rm MSP} = (0.39-1.08) \times \eta_{\rm Geminga}$. This conclusion is also supported by a flattening in the observed correlation between the far-infrared and radio luminosities of old star-forming galaxies~\cite{Sudoh:2020hyu}. This result has important implications for the origin of Galactic Center gamma-ray excess, as it indicates that if MSPs are responsible for the excess observed by Fermi, then they should also produce TeV-scale emission at a level that would exceed or saturate that observed from the Inner Galaxy by HESS. We look forward to observations by CTA, which should be able to either clearly identify the TeV halo emission from a large MSP population, or strongly constrain the number of MSPs that are present in the Inner Galaxy.

\bigskip

\section*{Acknowledgements}

\noindent DH is supported by the Fermi Research Alliance, LLC under Contract No. DE-AC02-07CH11359 with the U.S. Department of Energy, Office of High Energy Physics. TL is supported by the Swedish Research Council under contract 2019-05135, the Swedish National Space Agency under contract 117/19, and the European Research Council under grant 742104.

\bibliography{main}

\begin{thebibliography}{87}%
\makeatletter
\providecommand \@ifxundefined [1]{%
 \@ifx{#1\undefined}
}%
\providecommand \@ifnum [1]{%
 \ifnum #1\expandafter \@firstoftwo
 \else \expandafter \@secondoftwo
 \fi
}%
\providecommand \@ifx [1]{%
 \ifx #1\expandafter \@firstoftwo
 \else \expandafter \@secondoftwo
 \fi
}%
\providecommand \natexlab [1]{#1}%
\providecommand \enquote  [1]{``#1''}%
\providecommand \bibnamefont  [1]{#1}%
\providecommand \bibfnamefont [1]{#1}%
\providecommand \citenamefont [1]{#1}%
\providecommand \href@noop [0]{\@secondoftwo}%
\providecommand \href [0]{\begingroup \@sanitize@url \@href}%
\providecommand \@href[1]{\@@startlink{#1}\@@href}%
\providecommand \@@href[1]{\endgroup#1\@@endlink}%
\providecommand \@sanitize@url [0]{\catcode `\\12\catcode `\$12\catcode
  `\&12\catcode `\#12\catcode `\^12\catcode `\_12\catcode `\%12\relax}%
\providecommand \@@startlink[1]{}%
\providecommand \@@endlink[0]{}%
\providecommand \url  [0]{\begingroup\@sanitize@url \@url }%
\providecommand \@url [1]{\endgroup\@href {#1}{\urlprefix }}%
\providecommand \urlprefix  [0]{URL }%
\providecommand \Eprint [0]{\href }%
\providecommand \doibase [0]{http://dx.doi.org/}%
\providecommand \selectlanguage [0]{\@gobble}%
\providecommand \bibinfo  [0]{\@secondoftwo}%
\providecommand \bibfield  [0]{\@secondoftwo}%
\providecommand \translation [1]{[#1]}%
\providecommand \BibitemOpen [0]{}%
\providecommand \bibitemStop [0]{}%
\providecommand \bibitemNoStop [0]{.\EOS\space}%
\providecommand \EOS [0]{\spacefactor3000\relax}%
\providecommand \BibitemShut  [1]{\csname bibitem#1\endcsname}%
\let\auto@bib@innerbib\@empty
\bibitem [{\citenamefont {Albert}\ \emph {et~al.}(2020)\citenamefont {Albert}
  \emph {et~al.}}]{Albert:2020fua}%
  \BibitemOpen
  \bibfield  {author} {\bibinfo {author} {\bibfnamefont {A.}~\bibnamefont
  {Albert}} \emph {et~al.} (\bibinfo {collaboration} {HAWC}),\ }\href {\doibase
  10.3847/1538-4357/abc2d8} {\bibfield  {journal} {\bibinfo  {journal}
  {Astrophys. J.}\ }\textbf {\bibinfo {volume} {905}},\ \bibinfo {pages} {76}
  (\bibinfo {year} {2020})},\ \Eprint {http://arxiv.org/abs/2007.08582}
  {arXiv:2007.08582 [astro-ph.HE]} \BibitemShut {NoStop}%
\bibitem [{\citenamefont {Abeysekara}\ \emph
  {et~al.}(2017{\natexlab{a}})\citenamefont {Abeysekara} \emph
  {et~al.}}]{Abeysekara:2017hyn}%
  \BibitemOpen
  \bibfield  {author} {\bibinfo {author} {\bibfnamefont {A.}~\bibnamefont
  {Abeysekara}} \emph {et~al.},\ }\href {\doibase 10.3847/1538-4357/aa7556}
  {\bibfield  {journal} {\bibinfo  {journal} {Astrophys. J.}\ }\textbf
  {\bibinfo {volume} {843}},\ \bibinfo {pages} {40} (\bibinfo {year}
  {2017}{\natexlab{a}})},\ \Eprint {http://arxiv.org/abs/1702.02992}
  {arXiv:1702.02992 [astro-ph.HE]} \BibitemShut {NoStop}%
\bibitem [{\citenamefont {Abeysekara}\ \emph
  {et~al.}(2017{\natexlab{b}})\citenamefont {Abeysekara} \emph
  {et~al.}}]{Abeysekara:2017old}%
  \BibitemOpen
  \bibfield  {author} {\bibinfo {author} {\bibfnamefont {A.}~\bibnamefont
  {Abeysekara}} \emph {et~al.} (\bibinfo {collaboration} {HAWC}),\ }\href
  {\doibase 10.1126/science.aan4880} {\bibfield  {journal} {\bibinfo  {journal}
  {Science}\ }\textbf {\bibinfo {volume} {358}},\ \bibinfo {pages} {911}
  (\bibinfo {year} {2017}{\natexlab{b}})},\ \Eprint
  {http://arxiv.org/abs/1711.06223} {arXiv:1711.06223 [astro-ph.HE]}
  \BibitemShut {NoStop}%
\bibitem [{\citenamefont {{Abdo}}\ \emph {et~al.}(2009)\citenamefont {{Abdo}}
  \emph {et~al.}}]{Abdo:2009ku}%
  \BibitemOpen
  \bibfield  {author} {\bibinfo {author} {\bibfnamefont {A.~A.}\ \bibnamefont
  {{Abdo}}} \emph {et~al.},\ }\href {\doibase 10.1088/0004-637X/700/2/L127}
  {\bibfield  {journal} {\bibinfo  {journal} {The Astrophysical Journal
  Letters}\ }\textbf {\bibinfo {volume} {700}},\ \bibinfo {pages} {L127}
  (\bibinfo {year} {2009})},\ \Eprint {http://arxiv.org/abs/0904.1018}
  {arXiv:0904.1018 [astro-ph.HE]} \BibitemShut {NoStop}%
\bibitem [{\citenamefont {Hooper}\ \emph {et~al.}(2017)\citenamefont {Hooper},
  \citenamefont {Cholis}, \citenamefont {Linden},\ and\ \citenamefont
  {Fang}}]{Hooper:2017gtd}%
  \BibitemOpen
  \bibfield  {author} {\bibinfo {author} {\bibfnamefont {D.}~\bibnamefont
  {Hooper}}, \bibinfo {author} {\bibfnamefont {I.}~\bibnamefont {Cholis}},
  \bibinfo {author} {\bibfnamefont {T.}~\bibnamefont {Linden}}, \ and\ \bibinfo
  {author} {\bibfnamefont {K.}~\bibnamefont {Fang}},\ }\href {\doibase
  10.1103/PhysRevD.96.103013} {\bibfield  {journal} {\bibinfo  {journal} {Phys.
  Rev. D}\ }\textbf {\bibinfo {volume} {96}},\ \bibinfo {pages} {103013}
  (\bibinfo {year} {2017})},\ \Eprint {http://arxiv.org/abs/1702.08436}
  {arXiv:1702.08436 [astro-ph.HE]} \BibitemShut {NoStop}%
\bibitem [{\citenamefont {Hooper}\ and\ \citenamefont
  {Linden}(2018{\natexlab{a}})}]{Hooper:2017tkg}%
  \BibitemOpen
  \bibfield  {author} {\bibinfo {author} {\bibfnamefont {D.}~\bibnamefont
  {Hooper}}\ and\ \bibinfo {author} {\bibfnamefont {T.}~\bibnamefont
  {Linden}},\ }\href {\doibase 10.1103/PhysRevD.98.083009} {\bibfield
  {journal} {\bibinfo  {journal} {Phys. Rev. D}\ }\textbf {\bibinfo {volume}
  {98}},\ \bibinfo {pages} {083009} (\bibinfo {year} {2018}{\natexlab{a}})},\
  \Eprint {http://arxiv.org/abs/1711.07482} {arXiv:1711.07482 [astro-ph.HE]}
  \BibitemShut {NoStop}%
\bibitem [{\citenamefont {Johannesson}\ \emph {et~al.}(2019)\citenamefont
  {Johannesson}, \citenamefont {Porter},\ and\ \citenamefont
  {Moskalenko}}]{Johannesson:2019jlk}%
  \BibitemOpen
  \bibfield  {author} {\bibinfo {author} {\bibfnamefont {G.}~\bibnamefont
  {Johannesson}}, \bibinfo {author} {\bibfnamefont {T.~A.}\ \bibnamefont
  {Porter}}, \ and\ \bibinfo {author} {\bibfnamefont {I.~V.}\ \bibnamefont
  {Moskalenko}},\ }\href {\doibase 10.3847/1538-4357/ab258e} {\bibfield
  {journal} {\bibinfo  {journal} {Astrophys. J.}\ }\textbf {\bibinfo {volume}
  {879}},\ \bibinfo {pages} {91} (\bibinfo {year} {2019})},\ \Eprint
  {http://arxiv.org/abs/1903.05509} {arXiv:1903.05509 [astro-ph.HE]}
  \BibitemShut {NoStop}%
\bibitem [{\citenamefont {Di~Mauro}\ \emph {et~al.}(2020)\citenamefont
  {Di~Mauro}, \citenamefont {Manconi},\ and\ \citenamefont
  {Donato}}]{DiMauro:2019hwn}%
  \BibitemOpen
  \bibfield  {author} {\bibinfo {author} {\bibfnamefont {M.}~\bibnamefont
  {Di~Mauro}}, \bibinfo {author} {\bibfnamefont {S.}~\bibnamefont {Manconi}}, \
  and\ \bibinfo {author} {\bibfnamefont {F.}~\bibnamefont {Donato}},\ }\href
  {\doibase 10.1103/PhysRevD.101.103035} {\bibfield  {journal} {\bibinfo
  {journal} {Phys. Rev. D}\ }\textbf {\bibinfo {volume} {101}},\ \bibinfo
  {pages} {103035} (\bibinfo {year} {2020})},\ \Eprint
  {http://arxiv.org/abs/1908.03216} {arXiv:1908.03216 [astro-ph.HE]}
  \BibitemShut {NoStop}%
\bibitem [{\citenamefont {Liu}\ \emph {et~al.}(2019)\citenamefont {Liu},
  \citenamefont {Yan},\ and\ \citenamefont {Zhang}}]{Liu:2019zyj}%
  \BibitemOpen
  \bibfield  {author} {\bibinfo {author} {\bibfnamefont {R.-Y.}\ \bibnamefont
  {Liu}}, \bibinfo {author} {\bibfnamefont {H.}~\bibnamefont {Yan}}, \ and\
  \bibinfo {author} {\bibfnamefont {H.}~\bibnamefont {Zhang}},\ }\href
  {\doibase 10.1103/PhysRevLett.123.221103} {\bibfield  {journal} {\bibinfo
  {journal} {Phys. Rev. Lett.}\ }\textbf {\bibinfo {volume} {123}},\ \bibinfo
  {pages} {221103} (\bibinfo {year} {2019})},\ \Eprint
  {http://arxiv.org/abs/1904.11536} {arXiv:1904.11536 [astro-ph.HE]}
  \BibitemShut {NoStop}%
\bibitem [{\citenamefont {Smith}(2020)}]{Smith:2020clm}%
  \BibitemOpen
  \bibfield  {author} {\bibinfo {author} {\bibfnamefont {A.}~\bibnamefont
  {Smith}} (\bibinfo {collaboration} {HAWC}),\ }\href {\doibase
  10.22323/1.358.0797} {\bibfield  {journal} {\bibinfo  {journal} {PoS}\
  }\textbf {\bibinfo {volume} {ICRC2019}},\ \bibinfo {pages} {797} (\bibinfo
  {year} {2020})}\BibitemShut {NoStop}%
\bibitem [{\citenamefont {Albert}\ \emph {et~al.}(2021)\citenamefont {Albert}
  \emph {et~al.}}]{Albert:2021vrd}%
  \BibitemOpen
  \bibfield  {author} {\bibinfo {author} {\bibfnamefont {A.}~\bibnamefont
  {Albert}} \emph {et~al.} (\bibinfo {collaboration} {HAWC}),\ }\href@noop {}
  {\  (\bibinfo {year} {2021})},\ \Eprint {http://arxiv.org/abs/2101.07895}
  {arXiv:2101.07895 [astro-ph.HE]} \BibitemShut {NoStop}%
\bibitem [{\citenamefont {Abdalla}\ \emph
  {et~al.}(2018{\natexlab{a}})\citenamefont {Abdalla} \emph
  {et~al.}}]{H.E.S.S.:2018zkf}%
  \BibitemOpen
  \bibfield  {author} {\bibinfo {author} {\bibfnamefont {H.}~\bibnamefont
  {Abdalla}} \emph {et~al.} (\bibinfo {collaboration} {HESS}),\ }\href
  {\doibase 10.1051/0004-6361/201732098} {\bibfield  {journal} {\bibinfo
  {journal} {Astron. Astrophys.}\ }\textbf {\bibinfo {volume} {612}},\ \bibinfo
  {pages} {A1} (\bibinfo {year} {2018}{\natexlab{a}})},\ \Eprint
  {http://arxiv.org/abs/1804.02432} {arXiv:1804.02432 [astro-ph.HE]}
  \BibitemShut {NoStop}%
\bibitem [{\citenamefont {Abdalla}\ \emph
  {et~al.}(2018{\natexlab{b}})\citenamefont {Abdalla} \emph
  {et~al.}}]{Abdalla:2017vci}%
  \BibitemOpen
  \bibfield  {author} {\bibinfo {author} {\bibfnamefont {H.}~\bibnamefont
  {Abdalla}} \emph {et~al.} (\bibinfo {collaboration} {HESS}),\ }\href
  {\doibase 10.1051/0004-6361/201629377} {\bibfield  {journal} {\bibinfo
  {journal} {Astron. Astrophys.}\ }\textbf {\bibinfo {volume} {612}},\ \bibinfo
  {pages} {A2} (\bibinfo {year} {2018}{\natexlab{b}})},\ \Eprint
  {http://arxiv.org/abs/1702.08280} {arXiv:1702.08280 [astro-ph.HE]}
  \BibitemShut {NoStop}%
\bibitem [{\citenamefont {Abeysekara}\ \emph {et~al.}(2020)\citenamefont
  {Abeysekara} \emph {et~al.}}]{Abeysekara:2019gov}%
  \BibitemOpen
  \bibfield  {author} {\bibinfo {author} {\bibfnamefont {A.}~\bibnamefont
  {Abeysekara}} \emph {et~al.} (\bibinfo {collaboration} {HAWC}),\ }\href
  {\doibase 10.1103/PhysRevLett.124.021102} {\bibfield  {journal} {\bibinfo
  {journal} {Phys. Rev. Lett.}\ }\textbf {\bibinfo {volume} {124}},\ \bibinfo
  {pages} {021102} (\bibinfo {year} {2020})},\ \Eprint
  {http://arxiv.org/abs/1909.08609} {arXiv:1909.08609 [astro-ph.HE]}
  \BibitemShut {NoStop}%
\bibitem [{\citenamefont {Linden}\ \emph {et~al.}(2017)\citenamefont {Linden},
  \citenamefont {Auchettl}, \citenamefont {Bramante}, \citenamefont {Cholis},
  \citenamefont {Fang}, \citenamefont {Hooper}, \citenamefont {Karwal},\ and\
  \citenamefont {Li}}]{Linden:2017vvb}%
  \BibitemOpen
  \bibfield  {author} {\bibinfo {author} {\bibfnamefont {T.}~\bibnamefont
  {Linden}}, \bibinfo {author} {\bibfnamefont {K.}~\bibnamefont {Auchettl}},
  \bibinfo {author} {\bibfnamefont {J.}~\bibnamefont {Bramante}}, \bibinfo
  {author} {\bibfnamefont {I.}~\bibnamefont {Cholis}}, \bibinfo {author}
  {\bibfnamefont {K.}~\bibnamefont {Fang}}, \bibinfo {author} {\bibfnamefont
  {D.}~\bibnamefont {Hooper}}, \bibinfo {author} {\bibfnamefont
  {T.}~\bibnamefont {Karwal}}, \ and\ \bibinfo {author} {\bibfnamefont {S.~W.}\
  \bibnamefont {Li}},\ }\href {\doibase 10.1103/PhysRevD.96.103016} {\bibfield
  {journal} {\bibinfo  {journal} {Phys. Rev. D}\ }\textbf {\bibinfo {volume}
  {96}},\ \bibinfo {pages} {103016} (\bibinfo {year} {2017})},\ \Eprint
  {http://arxiv.org/abs/1703.09704} {arXiv:1703.09704 [astro-ph.HE]}
  \BibitemShut {NoStop}%
\bibitem [{\citenamefont {Sudoh}\ \emph {et~al.}(2019)\citenamefont {Sudoh},
  \citenamefont {Linden},\ and\ \citenamefont {Beacom}}]{Sudoh:2019lav}%
  \BibitemOpen
  \bibfield  {author} {\bibinfo {author} {\bibfnamefont {T.}~\bibnamefont
  {Sudoh}}, \bibinfo {author} {\bibfnamefont {T.}~\bibnamefont {Linden}}, \
  and\ \bibinfo {author} {\bibfnamefont {J.~F.}\ \bibnamefont {Beacom}},\
  }\href {\doibase 10.1103/PhysRevD.100.043016} {\bibfield  {journal} {\bibinfo
   {journal} {Phys. Rev. D}\ }\textbf {\bibinfo {volume} {100}},\ \bibinfo
  {pages} {043016} (\bibinfo {year} {2019})},\ \Eprint
  {http://arxiv.org/abs/1902.08203} {arXiv:1902.08203 [astro-ph.HE]}
  \BibitemShut {NoStop}%
\bibitem [{\citenamefont {Sudoh}\ \emph {et~al.}(2021)\citenamefont {Sudoh},
  \citenamefont {Linden},\ and\ \citenamefont {Hooper}}]{Sudoh:2021avj}%
  \BibitemOpen
  \bibfield  {author} {\bibinfo {author} {\bibfnamefont {T.}~\bibnamefont
  {Sudoh}}, \bibinfo {author} {\bibfnamefont {T.}~\bibnamefont {Linden}}, \
  and\ \bibinfo {author} {\bibfnamefont {D.}~\bibnamefont {Hooper}},\
  }\href@noop {} {\  (\bibinfo {year} {2021})},\ \Eprint
  {http://arxiv.org/abs/2101.11026} {arXiv:2101.11026 [astro-ph.HE]}
  \BibitemShut {NoStop}%
\bibitem [{\citenamefont {Venter}\ \emph {et~al.}(2015)\citenamefont {Venter},
  \citenamefont {Kopp}, \citenamefont {Harding}, \citenamefont {Gonthier},\
  and\ \citenamefont {B\"usching}}]{Venter:2015gga}%
  \BibitemOpen
  \bibfield  {author} {\bibinfo {author} {\bibfnamefont {C.}~\bibnamefont
  {Venter}}, \bibinfo {author} {\bibfnamefont {A.}~\bibnamefont {Kopp}},
  \bibinfo {author} {\bibfnamefont {A.~K.}\ \bibnamefont {Harding}}, \bibinfo
  {author} {\bibfnamefont {P.~L.}\ \bibnamefont {Gonthier}}, \ and\ \bibinfo
  {author} {\bibfnamefont {I.}~\bibnamefont {B\"usching}},\ }\href {\doibase
  10.1088/0004-637X/807/2/130} {\bibfield  {journal} {\bibinfo  {journal}
  {Astrophys. J.}\ }\textbf {\bibinfo {volume} {807}},\ \bibinfo {pages} {130}
  (\bibinfo {year} {2015})},\ \Eprint {http://arxiv.org/abs/1506.01211}
  {arXiv:1506.01211 [astro-ph.HE]} \BibitemShut {NoStop}%
\bibitem [{\citenamefont {Bednarek}\ \emph {et~al.}(2016)\citenamefont
  {Bednarek}, \citenamefont {Sitarek},\ and\ \citenamefont
  {Sobczak}}]{Bednarek:2016gpp}%
  \BibitemOpen
  \bibfield  {author} {\bibinfo {author} {\bibfnamefont {W.}~\bibnamefont
  {Bednarek}}, \bibinfo {author} {\bibfnamefont {J.}~\bibnamefont {Sitarek}}, \
  and\ \bibinfo {author} {\bibfnamefont {T.}~\bibnamefont {Sobczak}},\ }\href
  {\doibase 10.1093/mnras/stw367} {\bibfield  {journal} {\bibinfo  {journal}
  {Mon. Not. Roy. Astron. Soc.}\ }\textbf {\bibinfo {volume} {458}},\ \bibinfo
  {pages} {1083} (\bibinfo {year} {2016})},\ \Eprint
  {http://arxiv.org/abs/1602.03629} {arXiv:1602.03629 [astro-ph.HE]}
  \BibitemShut {NoStop}%
\bibitem [{\citenamefont {Venter}\ \emph {et~al.}(2016)\citenamefont {Venter},
  \citenamefont {Kopp}, \citenamefont {Harding}, \citenamefont {Gonthier},\
  and\ \citenamefont {B\"usching}}]{Venter:2015oza}%
  \BibitemOpen
  \bibfield  {author} {\bibinfo {author} {\bibfnamefont {C.}~\bibnamefont
  {Venter}}, \bibinfo {author} {\bibfnamefont {A.}~\bibnamefont {Kopp}},
  \bibinfo {author} {\bibfnamefont {A.~K.}\ \bibnamefont {Harding}}, \bibinfo
  {author} {\bibfnamefont {P.~L.}\ \bibnamefont {Gonthier}}, \ and\ \bibinfo
  {author} {\bibfnamefont {I.}~\bibnamefont {B\"usching}},\ }\href {\doibase
  10.22323/1.236.0462} {\bibfield  {journal} {\bibinfo  {journal} {PoS}\
  }\textbf {\bibinfo {volume} {ICRC2015}},\ \bibinfo {pages} {462} (\bibinfo
  {year} {2016})},\ \Eprint {http://arxiv.org/abs/1508.04676} {arXiv:1508.04676
  [astro-ph.HE]} \BibitemShut {NoStop}%
\bibitem [{\citenamefont {{Sironi}}\ and\ \citenamefont
  {{Spitkovsky}}(2011)}]{2011ApJ...741...39S}%
  \BibitemOpen
  \bibfield  {author} {\bibinfo {author} {\bibfnamefont {L.}~\bibnamefont
  {{Sironi}}}\ and\ \bibinfo {author} {\bibfnamefont {A.}~\bibnamefont
  {{Spitkovsky}}},\ }\href {\doibase 10.1088/0004-637X/741/1/39} {\bibfield
  {journal} {\bibinfo  {journal} {\apj}\ }\textbf {\bibinfo {volume} {741}},\
  \bibinfo {eid} {39} (\bibinfo {year} {2011})},\ \Eprint
  {http://arxiv.org/abs/1107.0977} {arXiv:1107.0977 [astro-ph.HE]} \BibitemShut
  {NoStop}%
\bibitem [{\citenamefont {Gaensler}\ and\ \citenamefont
  {Slane}(2006)}]{Gaensler:2006ua}%
  \BibitemOpen
  \bibfield  {author} {\bibinfo {author} {\bibfnamefont {B.~M.}\ \bibnamefont
  {Gaensler}}\ and\ \bibinfo {author} {\bibfnamefont {P.~O.}\ \bibnamefont
  {Slane}},\ }\href {\doibase 10.1146/annurev.astro.44.051905.092528}
  {\bibfield  {journal} {\bibinfo  {journal} {Ann. Rev. Astron. Astrophys.}\
  }\textbf {\bibinfo {volume} {44}},\ \bibinfo {pages} {17} (\bibinfo {year}
  {2006})},\ \Eprint {http://arxiv.org/abs/astro-ph/0601081}
  {arXiv:astro-ph/0601081} \BibitemShut {NoStop}%
\bibitem [{\citenamefont {Evoli}\ \emph {et~al.}(2018)\citenamefont {Evoli},
  \citenamefont {Linden},\ and\ \citenamefont {Morlino}}]{Evoli:2018aza}%
  \BibitemOpen
  \bibfield  {author} {\bibinfo {author} {\bibfnamefont {C.}~\bibnamefont
  {Evoli}}, \bibinfo {author} {\bibfnamefont {T.}~\bibnamefont {Linden}}, \
  and\ \bibinfo {author} {\bibfnamefont {G.}~\bibnamefont {Morlino}},\ }\href
  {\doibase 10.1103/PhysRevD.98.063017} {\bibfield  {journal} {\bibinfo
  {journal} {Phys. Rev. D}\ }\textbf {\bibinfo {volume} {98}},\ \bibinfo
  {pages} {063017} (\bibinfo {year} {2018})},\ \Eprint
  {http://arxiv.org/abs/1807.09263} {arXiv:1807.09263 [astro-ph.HE]}
  \BibitemShut {NoStop}%
\bibitem [{\citenamefont {Fang}\ \emph {et~al.}(2019)\citenamefont {Fang},
  \citenamefont {Bi},\ and\ \citenamefont {Yin}}]{Kun:2019sks}%
  \BibitemOpen
  \bibfield  {author} {\bibinfo {author} {\bibfnamefont {K.}~\bibnamefont
  {Fang}}, \bibinfo {author} {\bibfnamefont {X.-J.}\ \bibnamefont {Bi}}, \ and\
  \bibinfo {author} {\bibfnamefont {P.-F.}\ \bibnamefont {Yin}},\ }\href
  {\doibase 10.1093/mnras/stz1974} {\bibfield  {journal} {\bibinfo  {journal}
  {Mon. Not. Roy. Astron. Soc.}\ }\textbf {\bibinfo {volume} {488}},\ \bibinfo
  {pages} {4074} (\bibinfo {year} {2019})},\ \Eprint
  {http://arxiv.org/abs/1903.06421} {arXiv:1903.06421 [astro-ph.HE]}
  \BibitemShut {NoStop}%
\bibitem [{\citenamefont {Hooper}\ and\ \citenamefont
  {Goodenough}(2011)}]{Hooper:2010mq}%
  \BibitemOpen
  \bibfield  {author} {\bibinfo {author} {\bibfnamefont {D.}~\bibnamefont
  {Hooper}}\ and\ \bibinfo {author} {\bibfnamefont {L.}~\bibnamefont
  {Goodenough}},\ }\href {\doibase 10.1016/j.physletb.2011.02.029} {\bibfield
  {journal} {\bibinfo  {journal} {Phys. Lett. B}\ }\textbf {\bibinfo {volume}
  {697}},\ \bibinfo {pages} {412} (\bibinfo {year} {2011})},\ \Eprint
  {http://arxiv.org/abs/1010.2752} {arXiv:1010.2752 [hep-ph]} \BibitemShut
  {NoStop}%
\bibitem [{\citenamefont {Daylan}\ \emph {et~al.}(2016)\citenamefont {Daylan},
  \citenamefont {Finkbeiner}, \citenamefont {Hooper}, \citenamefont {Linden},
  \citenamefont {Portillo}, \citenamefont {Rodd},\ and\ \citenamefont
  {Slatyer}}]{Daylan:2014rsa}%
  \BibitemOpen
  \bibfield  {author} {\bibinfo {author} {\bibfnamefont {T.}~\bibnamefont
  {Daylan}}, \bibinfo {author} {\bibfnamefont {D.~P.}\ \bibnamefont
  {Finkbeiner}}, \bibinfo {author} {\bibfnamefont {D.}~\bibnamefont {Hooper}},
  \bibinfo {author} {\bibfnamefont {T.}~\bibnamefont {Linden}}, \bibinfo
  {author} {\bibfnamefont {S.~K.~N.}\ \bibnamefont {Portillo}}, \bibinfo
  {author} {\bibfnamefont {N.~L.}\ \bibnamefont {Rodd}}, \ and\ \bibinfo
  {author} {\bibfnamefont {T.~R.}\ \bibnamefont {Slatyer}},\ }\href {\doibase
  10.1016/j.dark.2015.12.005} {\bibfield  {journal} {\bibinfo  {journal} {Phys.
  Dark Univ.}\ }\textbf {\bibinfo {volume} {12}},\ \bibinfo {pages} {1}
  (\bibinfo {year} {2016})},\ \Eprint {http://arxiv.org/abs/1402.6703}
  {arXiv:1402.6703 [astro-ph.HE]} \BibitemShut {NoStop}%
\bibitem [{\citenamefont {Abazajian}(2011)}]{Abazajian:2010zy}%
  \BibitemOpen
  \bibfield  {author} {\bibinfo {author} {\bibfnamefont {K.~N.}\ \bibnamefont
  {Abazajian}},\ }\href {\doibase 10.1088/1475-7516/2011/03/010} {\bibfield
  {journal} {\bibinfo  {journal} {JCAP}\ }\textbf {\bibinfo {volume} {03}},\
  \bibinfo {pages} {010} (\bibinfo {year} {2011})},\ \Eprint
  {http://arxiv.org/abs/1011.4275} {arXiv:1011.4275 [astro-ph.HE]} \BibitemShut
  {NoStop}%
\bibitem [{\citenamefont {Lee}\ \emph {et~al.}(2016)\citenamefont {Lee},
  \citenamefont {Lisanti}, \citenamefont {Safdi}, \citenamefont {Slatyer},\
  and\ \citenamefont {Xue}}]{Lee:2015fea}%
  \BibitemOpen
  \bibfield  {author} {\bibinfo {author} {\bibfnamefont {S.~K.}\ \bibnamefont
  {Lee}}, \bibinfo {author} {\bibfnamefont {M.}~\bibnamefont {Lisanti}},
  \bibinfo {author} {\bibfnamefont {B.~R.}\ \bibnamefont {Safdi}}, \bibinfo
  {author} {\bibfnamefont {T.~R.}\ \bibnamefont {Slatyer}}, \ and\ \bibinfo
  {author} {\bibfnamefont {W.}~\bibnamefont {Xue}},\ }\href {\doibase
  10.1103/PhysRevLett.116.051103} {\bibfield  {journal} {\bibinfo  {journal}
  {Phys. Rev. Lett.}\ }\textbf {\bibinfo {volume} {116}},\ \bibinfo {pages}
  {051103} (\bibinfo {year} {2016})},\ \Eprint
  {http://arxiv.org/abs/1506.05124} {arXiv:1506.05124 [astro-ph.HE]}
  \BibitemShut {NoStop}%
\bibitem [{\citenamefont {Bartels}\ \emph {et~al.}(2016)\citenamefont
  {Bartels}, \citenamefont {Krishnamurthy},\ and\ \citenamefont
  {Weniger}}]{Bartels:2015aea}%
  \BibitemOpen
  \bibfield  {author} {\bibinfo {author} {\bibfnamefont {R.}~\bibnamefont
  {Bartels}}, \bibinfo {author} {\bibfnamefont {S.}~\bibnamefont
  {Krishnamurthy}}, \ and\ \bibinfo {author} {\bibfnamefont {C.}~\bibnamefont
  {Weniger}},\ }\href {\doibase 10.1103/PhysRevLett.116.051102} {\bibfield
  {journal} {\bibinfo  {journal} {Phys. Rev. Lett.}\ }\textbf {\bibinfo
  {volume} {116}},\ \bibinfo {pages} {051102} (\bibinfo {year} {2016})},\
  \Eprint {http://arxiv.org/abs/1506.05104} {arXiv:1506.05104 [astro-ph.HE]}
  \BibitemShut {NoStop}%
\bibitem [{\citenamefont {Hooper}\ and\ \citenamefont
  {Linden}(2018{\natexlab{b}})}]{Hooper:2018fih}%
  \BibitemOpen
  \bibfield  {author} {\bibinfo {author} {\bibfnamefont {D.}~\bibnamefont
  {Hooper}}\ and\ \bibinfo {author} {\bibfnamefont {T.}~\bibnamefont
  {Linden}},\ }\href {\doibase 10.1103/PhysRevD.98.043005} {\bibfield
  {journal} {\bibinfo  {journal} {Phys. Rev. D}\ }\textbf {\bibinfo {volume}
  {98}},\ \bibinfo {pages} {043005} (\bibinfo {year} {2018}{\natexlab{b}})},\
  \Eprint {http://arxiv.org/abs/1803.08046} {arXiv:1803.08046 [astro-ph.HE]}
  \BibitemShut {NoStop}%
\bibitem [{\citenamefont {Manchester}\ \emph {et~al.}(2005)\citenamefont
  {Manchester}, \citenamefont {Hobbs}, \citenamefont {Teoh},\ and\
  \citenamefont {Hobbs}}]{Manchester:2004bp}%
  \BibitemOpen
  \bibfield  {author} {\bibinfo {author} {\bibfnamefont {R.~N.}\ \bibnamefont
  {Manchester}}, \bibinfo {author} {\bibfnamefont {G.~B.}\ \bibnamefont
  {Hobbs}}, \bibinfo {author} {\bibfnamefont {A.}~\bibnamefont {Teoh}}, \ and\
  \bibinfo {author} {\bibfnamefont {M.}~\bibnamefont {Hobbs}},\ }\href
  {\doibase 10.1086/428488} {\bibfield  {journal} {\bibinfo  {journal} {Astron.
  J.}\ }\textbf {\bibinfo {volume} {129}},\ \bibinfo {pages} {1993} (\bibinfo
  {year} {2005})},\ \Eprint {http://arxiv.org/abs/astro-ph/0412641}
  {arXiv:astro-ph/0412641} \BibitemShut {NoStop}%
\bibitem [{\citenamefont {Ackermann}\ \emph {et~al.}(2017)\citenamefont
  {Ackermann} \emph {et~al.}}]{TheFermi-LAT:2017vmf}%
  \BibitemOpen
  \bibfield  {author} {\bibinfo {author} {\bibfnamefont {M.}~\bibnamefont
  {Ackermann}} \emph {et~al.} (\bibinfo {collaboration} {Fermi-LAT}),\ }\href
  {\doibase 10.3847/1538-4357/aa6cab} {\bibfield  {journal} {\bibinfo
  {journal} {Astrophys. J.}\ }\textbf {\bibinfo {volume} {840}},\ \bibinfo
  {pages} {43} (\bibinfo {year} {2017})},\ \Eprint
  {http://arxiv.org/abs/1704.03910} {arXiv:1704.03910 [astro-ph.HE]}
  \BibitemShut {NoStop}%
\bibitem [{\citenamefont {Goodenough}\ and\ \citenamefont
  {Hooper}(2009)}]{Goodenough:2009gk}%
  \BibitemOpen
  \bibfield  {author} {\bibinfo {author} {\bibfnamefont {L.}~\bibnamefont
  {Goodenough}}\ and\ \bibinfo {author} {\bibfnamefont {D.}~\bibnamefont
  {Hooper}},\ }\href@noop {} {\  (\bibinfo {year} {2009})},\ \Eprint
  {http://arxiv.org/abs/0910.2998} {arXiv:0910.2998 [hep-ph]} \BibitemShut
  {NoStop}%
\bibitem [{\citenamefont {Hooper}\ and\ \citenamefont
  {Linden}(2011)}]{Hooper:2011ti}%
  \BibitemOpen
  \bibfield  {author} {\bibinfo {author} {\bibfnamefont {D.}~\bibnamefont
  {Hooper}}\ and\ \bibinfo {author} {\bibfnamefont {T.}~\bibnamefont
  {Linden}},\ }\href {\doibase 10.1103/PhysRevD.84.123005} {\bibfield
  {journal} {\bibinfo  {journal} {Phys.Rev.}\ }\textbf {\bibinfo {volume}
  {D84}},\ \bibinfo {pages} {123005} (\bibinfo {year} {2011})},\ \Eprint
  {http://arxiv.org/abs/1110.0006} {arXiv:1110.0006 [astro-ph.HE]} \BibitemShut
  {NoStop}%
\bibitem [{\citenamefont {Abazajian}\ and\ \citenamefont
  {Kaplinghat}(2012)}]{Abazajian:2012pn}%
  \BibitemOpen
  \bibfield  {author} {\bibinfo {author} {\bibfnamefont {K.~N.}\ \bibnamefont
  {Abazajian}}\ and\ \bibinfo {author} {\bibfnamefont {M.}~\bibnamefont
  {Kaplinghat}},\ }\href {\doibase 10.1103/PhysRevD.86.083511} {\bibfield
  {journal} {\bibinfo  {journal} {Phys. Rev. D}\ }\textbf {\bibinfo {volume}
  {86}},\ \bibinfo {pages} {083511} (\bibinfo {year} {2012})},\ \bibinfo {note}
  {[Erratum: Phys.Rev.D 87, 129902 (2013)]},\ \Eprint
  {http://arxiv.org/abs/1207.6047} {arXiv:1207.6047 [astro-ph.HE]} \BibitemShut
  {NoStop}%
\bibitem [{\citenamefont {Hooper}\ and\ \citenamefont
  {Slatyer}(2013)}]{Hooper:2013rwa}%
  \BibitemOpen
  \bibfield  {author} {\bibinfo {author} {\bibfnamefont {D.}~\bibnamefont
  {Hooper}}\ and\ \bibinfo {author} {\bibfnamefont {T.~R.}\ \bibnamefont
  {Slatyer}},\ }\href {\doibase 10.1016/j.dark.2013.06.003} {\bibfield
  {journal} {\bibinfo  {journal} {Phys. Dark Univ.}\ }\textbf {\bibinfo
  {volume} {2}},\ \bibinfo {pages} {118} (\bibinfo {year} {2013})},\ \Eprint
  {http://arxiv.org/abs/1302.6589} {arXiv:1302.6589 [astro-ph.HE]} \BibitemShut
  {NoStop}%
\bibitem [{\citenamefont {Gordon}\ and\ \citenamefont
  {Macias}(2013)}]{Gordon:2013vta}%
  \BibitemOpen
  \bibfield  {author} {\bibinfo {author} {\bibfnamefont {C.}~\bibnamefont
  {Gordon}}\ and\ \bibinfo {author} {\bibfnamefont {O.}~\bibnamefont
  {Macias}},\ }\href {\doibase 10.1103/PhysRevD.88.083521} {\bibfield
  {journal} {\bibinfo  {journal} {Phys. Rev. D}\ }\textbf {\bibinfo {volume}
  {88}},\ \bibinfo {pages} {083521} (\bibinfo {year} {2013})},\ \bibinfo {note}
  {[Erratum: Phys.Rev.D 89, 049901 (2014)]},\ \Eprint
  {http://arxiv.org/abs/1306.5725} {arXiv:1306.5725 [astro-ph.HE]} \BibitemShut
  {NoStop}%
\bibitem [{\citenamefont {Calore}\ \emph {et~al.}(2015)\citenamefont {Calore},
  \citenamefont {Cholis},\ and\ \citenamefont {Weniger}}]{Calore:2014xka}%
  \BibitemOpen
  \bibfield  {author} {\bibinfo {author} {\bibfnamefont {F.}~\bibnamefont
  {Calore}}, \bibinfo {author} {\bibfnamefont {I.}~\bibnamefont {Cholis}}, \
  and\ \bibinfo {author} {\bibfnamefont {C.}~\bibnamefont {Weniger}},\ }\href
  {\doibase 10.1088/1475-7516/2015/03/038} {\bibfield  {journal} {\bibinfo
  {journal} {JCAP}\ }\textbf {\bibinfo {volume} {03}},\ \bibinfo {pages} {038}
  (\bibinfo {year} {2015})},\ \Eprint {http://arxiv.org/abs/1409.0042}
  {arXiv:1409.0042 [astro-ph.CO]} \BibitemShut {NoStop}%
\bibitem [{\citenamefont {Ajello}\ \emph {et~al.}(2016)\citenamefont {Ajello}
  \emph {et~al.}}]{TheFermi-LAT:2015kwa}%
  \BibitemOpen
  \bibfield  {author} {\bibinfo {author} {\bibfnamefont {M.}~\bibnamefont
  {Ajello}} \emph {et~al.} (\bibinfo {collaboration} {Fermi-LAT}),\ }\href
  {\doibase 10.3847/0004-637X/819/1/44} {\bibfield  {journal} {\bibinfo
  {journal} {Astrophys. J.}\ }\textbf {\bibinfo {volume} {819}},\ \bibinfo
  {pages} {44} (\bibinfo {year} {2016})},\ \Eprint
  {http://arxiv.org/abs/1511.02938} {arXiv:1511.02938 [astro-ph.HE]}
  \BibitemShut {NoStop}%
\bibitem [{\citenamefont {Berlin}\ \emph
  {et~al.}(2014{\natexlab{a}})\citenamefont {Berlin}, \citenamefont {Hooper},\
  and\ \citenamefont {McDermott}}]{Berlin:2014tja}%
  \BibitemOpen
  \bibfield  {author} {\bibinfo {author} {\bibfnamefont {A.}~\bibnamefont
  {Berlin}}, \bibinfo {author} {\bibfnamefont {D.}~\bibnamefont {Hooper}}, \
  and\ \bibinfo {author} {\bibfnamefont {S.~D.}\ \bibnamefont {McDermott}},\
  }\href {\doibase 10.1103/PhysRevD.89.115022} {\bibfield  {journal} {\bibinfo
  {journal} {Phys. Rev. D}\ }\textbf {\bibinfo {volume} {89}},\ \bibinfo
  {pages} {115022} (\bibinfo {year} {2014}{\natexlab{a}})},\ \Eprint
  {http://arxiv.org/abs/1404.0022} {arXiv:1404.0022 [hep-ph]} \BibitemShut
  {NoStop}%
\bibitem [{\citenamefont {Berlin}\ \emph
  {et~al.}(2014{\natexlab{b}})\citenamefont {Berlin}, \citenamefont {Gratia},
  \citenamefont {Hooper},\ and\ \citenamefont {McDermott}}]{Berlin:2014pya}%
  \BibitemOpen
  \bibfield  {author} {\bibinfo {author} {\bibfnamefont {A.}~\bibnamefont
  {Berlin}}, \bibinfo {author} {\bibfnamefont {P.}~\bibnamefont {Gratia}},
  \bibinfo {author} {\bibfnamefont {D.}~\bibnamefont {Hooper}}, \ and\ \bibinfo
  {author} {\bibfnamefont {S.~D.}\ \bibnamefont {McDermott}},\ }\href {\doibase
  10.1103/PhysRevD.90.015032} {\bibfield  {journal} {\bibinfo  {journal} {Phys.
  Rev. D}\ }\textbf {\bibinfo {volume} {90}},\ \bibinfo {pages} {015032}
  (\bibinfo {year} {2014}{\natexlab{b}})},\ \Eprint
  {http://arxiv.org/abs/1405.5204} {arXiv:1405.5204 [hep-ph]} \BibitemShut
  {NoStop}%
\bibitem [{\citenamefont {Abdullah}\ \emph {et~al.}(2014)\citenamefont
  {Abdullah}, \citenamefont {DiFranzo}, \citenamefont {Rajaraman},
  \citenamefont {Tait}, \citenamefont {Tanedo},\ and\ \citenamefont
  {Wijangco}}]{Abdullah:2014lla}%
  \BibitemOpen
  \bibfield  {author} {\bibinfo {author} {\bibfnamefont {M.}~\bibnamefont
  {Abdullah}}, \bibinfo {author} {\bibfnamefont {A.}~\bibnamefont {DiFranzo}},
  \bibinfo {author} {\bibfnamefont {A.}~\bibnamefont {Rajaraman}}, \bibinfo
  {author} {\bibfnamefont {T.~M.~P.}\ \bibnamefont {Tait}}, \bibinfo {author}
  {\bibfnamefont {P.}~\bibnamefont {Tanedo}}, \ and\ \bibinfo {author}
  {\bibfnamefont {A.~M.}\ \bibnamefont {Wijangco}},\ }\href {\doibase
  10.1103/PhysRevD.90.035004} {\bibfield  {journal} {\bibinfo  {journal} {Phys.
  Rev. D}\ }\textbf {\bibinfo {volume} {90}},\ \bibinfo {pages} {035004}
  (\bibinfo {year} {2014})},\ \Eprint {http://arxiv.org/abs/1404.6528}
  {arXiv:1404.6528 [hep-ph]} \BibitemShut {NoStop}%
\bibitem [{\citenamefont {Hooper}\ \emph {et~al.}(2012)\citenamefont {Hooper},
  \citenamefont {Weiner},\ and\ \citenamefont {Xue}}]{Hooper:2012cw}%
  \BibitemOpen
  \bibfield  {author} {\bibinfo {author} {\bibfnamefont {D.}~\bibnamefont
  {Hooper}}, \bibinfo {author} {\bibfnamefont {N.}~\bibnamefont {Weiner}}, \
  and\ \bibinfo {author} {\bibfnamefont {W.}~\bibnamefont {Xue}},\ }\href
  {\doibase 10.1103/PhysRevD.86.056009} {\bibfield  {journal} {\bibinfo
  {journal} {Phys. Rev. D}\ }\textbf {\bibinfo {volume} {86}},\ \bibinfo
  {pages} {056009} (\bibinfo {year} {2012})},\ \Eprint
  {http://arxiv.org/abs/1206.2929} {arXiv:1206.2929 [hep-ph]} \BibitemShut
  {NoStop}%
\bibitem [{\citenamefont {Martin}\ \emph {et~al.}(2014)\citenamefont {Martin},
  \citenamefont {Shelton},\ and\ \citenamefont {Unwin}}]{Martin:2014sxa}%
  \BibitemOpen
  \bibfield  {author} {\bibinfo {author} {\bibfnamefont {A.}~\bibnamefont
  {Martin}}, \bibinfo {author} {\bibfnamefont {J.}~\bibnamefont {Shelton}}, \
  and\ \bibinfo {author} {\bibfnamefont {J.}~\bibnamefont {Unwin}},\ }\href
  {\doibase 10.1103/PhysRevD.90.103513} {\bibfield  {journal} {\bibinfo
  {journal} {Phys. Rev. D}\ }\textbf {\bibinfo {volume} {90}},\ \bibinfo
  {pages} {103513} (\bibinfo {year} {2014})},\ \Eprint
  {http://arxiv.org/abs/1405.0272} {arXiv:1405.0272 [hep-ph]} \BibitemShut
  {NoStop}%
\bibitem [{\citenamefont {McDermott}(2015)}]{McDermott:2014rqa}%
  \BibitemOpen
  \bibfield  {author} {\bibinfo {author} {\bibfnamefont {S.~D.}\ \bibnamefont
  {McDermott}},\ }\href {\doibase 10.1016/j.dark.2015.05.001} {\bibfield
  {journal} {\bibinfo  {journal} {Phys. Dark Univ.}\ }\textbf {\bibinfo
  {volume} {7-8}},\ \bibinfo {pages} {12} (\bibinfo {year} {2015})},\ \Eprint
  {http://arxiv.org/abs/1406.6408} {arXiv:1406.6408 [hep-ph]} \BibitemShut
  {NoStop}%
\bibitem [{\citenamefont {Cahill-Rowley}\ \emph {et~al.}(2015)\citenamefont
  {Cahill-Rowley}, \citenamefont {Gainer}, \citenamefont {Hewett},\ and\
  \citenamefont {Rizzo}}]{Cahill-Rowley:2014ora}%
  \BibitemOpen
  \bibfield  {author} {\bibinfo {author} {\bibfnamefont {M.}~\bibnamefont
  {Cahill-Rowley}}, \bibinfo {author} {\bibfnamefont {J.}~\bibnamefont
  {Gainer}}, \bibinfo {author} {\bibfnamefont {J.}~\bibnamefont {Hewett}}, \
  and\ \bibinfo {author} {\bibfnamefont {T.}~\bibnamefont {Rizzo}},\ }\href
  {\doibase 10.1007/JHEP02(2015)057} {\bibfield  {journal} {\bibinfo  {journal}
  {JHEP}\ }\textbf {\bibinfo {volume} {02}},\ \bibinfo {pages} {057} (\bibinfo
  {year} {2015})},\ \Eprint {http://arxiv.org/abs/1409.1573} {arXiv:1409.1573
  [hep-ph]} \BibitemShut {NoStop}%
\bibitem [{\citenamefont {Hooper}(2015)}]{Hooper:2014fda}%
  \BibitemOpen
  \bibfield  {author} {\bibinfo {author} {\bibfnamefont {D.}~\bibnamefont
  {Hooper}},\ }\href {\doibase 10.1103/PhysRevD.91.035025} {\bibfield
  {journal} {\bibinfo  {journal} {Phys. Rev. D}\ }\textbf {\bibinfo {volume}
  {91}},\ \bibinfo {pages} {035025} (\bibinfo {year} {2015})},\ \Eprint
  {http://arxiv.org/abs/1411.4079} {arXiv:1411.4079 [hep-ph]} \BibitemShut
  {NoStop}%
\bibitem [{\citenamefont {Liu}\ \emph {et~al.}(2015)\citenamefont {Liu},
  \citenamefont {Weiner},\ and\ \citenamefont {Xue}}]{Liu:2014cma}%
  \BibitemOpen
  \bibfield  {author} {\bibinfo {author} {\bibfnamefont {J.}~\bibnamefont
  {Liu}}, \bibinfo {author} {\bibfnamefont {N.}~\bibnamefont {Weiner}}, \ and\
  \bibinfo {author} {\bibfnamefont {W.}~\bibnamefont {Xue}},\ }\href {\doibase
  10.1007/JHEP08(2015)050} {\bibfield  {journal} {\bibinfo  {journal} {JHEP}\
  }\textbf {\bibinfo {volume} {08}},\ \bibinfo {pages} {050} (\bibinfo {year}
  {2015})},\ \Eprint {http://arxiv.org/abs/1412.1485} {arXiv:1412.1485
  [hep-ph]} \BibitemShut {NoStop}%
\bibitem [{\citenamefont {Achterberg}\ \emph {et~al.}(2015)\citenamefont
  {Achterberg}, \citenamefont {Amoroso}, \citenamefont {Caron}, \citenamefont
  {Hendriks}, \citenamefont {Ruiz~de Austri},\ and\ \citenamefont
  {Weniger}}]{Caron:2015wda}%
  \BibitemOpen
  \bibfield  {author} {\bibinfo {author} {\bibfnamefont {A.}~\bibnamefont
  {Achterberg}}, \bibinfo {author} {\bibfnamefont {S.}~\bibnamefont {Amoroso}},
  \bibinfo {author} {\bibfnamefont {S.}~\bibnamefont {Caron}}, \bibinfo
  {author} {\bibfnamefont {L.}~\bibnamefont {Hendriks}}, \bibinfo {author}
  {\bibfnamefont {R.}~\bibnamefont {Ruiz~de Austri}}, \ and\ \bibinfo {author}
  {\bibfnamefont {C.}~\bibnamefont {Weniger}},\ }\href {\doibase
  10.1088/1475-7516/2015/08/006} {\bibfield  {journal} {\bibinfo  {journal}
  {JCAP}\ }\textbf {\bibinfo {volume} {08}},\ \bibinfo {pages} {006} (\bibinfo
  {year} {2015})},\ \Eprint {http://arxiv.org/abs/1502.05703} {arXiv:1502.05703
  [hep-ph]} \BibitemShut {NoStop}%
\bibitem [{\citenamefont {Berlin}\ \emph {et~al.}(2015)\citenamefont {Berlin},
  \citenamefont {Gori}, \citenamefont {Lin},\ and\ \citenamefont
  {Wang}}]{Berlin:2015wwa}%
  \BibitemOpen
  \bibfield  {author} {\bibinfo {author} {\bibfnamefont {A.}~\bibnamefont
  {Berlin}}, \bibinfo {author} {\bibfnamefont {S.}~\bibnamefont {Gori}},
  \bibinfo {author} {\bibfnamefont {T.}~\bibnamefont {Lin}}, \ and\ \bibinfo
  {author} {\bibfnamefont {L.-T.}\ \bibnamefont {Wang}},\ }\href {\doibase
  10.1103/PhysRevD.92.015005} {\bibfield  {journal} {\bibinfo  {journal} {Phys.
  Rev. D}\ }\textbf {\bibinfo {volume} {92}},\ \bibinfo {pages} {015005}
  (\bibinfo {year} {2015})},\ \Eprint {http://arxiv.org/abs/1502.06000}
  {arXiv:1502.06000 [hep-ph]} \BibitemShut {NoStop}%
\bibitem [{\citenamefont {Cline}\ \emph {et~al.}(2015)\citenamefont {Cline},
  \citenamefont {Dupuis}, \citenamefont {Liu},\ and\ \citenamefont
  {Xue}}]{Cline:2015qha}%
  \BibitemOpen
  \bibfield  {author} {\bibinfo {author} {\bibfnamefont {J.~M.}\ \bibnamefont
  {Cline}}, \bibinfo {author} {\bibfnamefont {G.}~\bibnamefont {Dupuis}},
  \bibinfo {author} {\bibfnamefont {Z.}~\bibnamefont {Liu}}, \ and\ \bibinfo
  {author} {\bibfnamefont {W.}~\bibnamefont {Xue}},\ }\href {\doibase
  10.1103/PhysRevD.91.115010} {\bibfield  {journal} {\bibinfo  {journal} {Phys.
  Rev. D}\ }\textbf {\bibinfo {volume} {91}},\ \bibinfo {pages} {115010}
  (\bibinfo {year} {2015})},\ \Eprint {http://arxiv.org/abs/1503.08213}
  {arXiv:1503.08213 [hep-ph]} \BibitemShut {NoStop}%
\bibitem [{\citenamefont {Bertone}\ \emph {et~al.}(2016)\citenamefont
  {Bertone}, \citenamefont {Calore}, \citenamefont {Caron}, \citenamefont
  {Ruiz}, \citenamefont {Kim}, \citenamefont {Trotta},\ and\ \citenamefont
  {Weniger}}]{Bertone:2015tza}%
  \BibitemOpen
  \bibfield  {author} {\bibinfo {author} {\bibfnamefont {G.}~\bibnamefont
  {Bertone}}, \bibinfo {author} {\bibfnamefont {F.}~\bibnamefont {Calore}},
  \bibinfo {author} {\bibfnamefont {S.}~\bibnamefont {Caron}}, \bibinfo
  {author} {\bibfnamefont {R.}~\bibnamefont {Ruiz}}, \bibinfo {author}
  {\bibfnamefont {J.~S.}\ \bibnamefont {Kim}}, \bibinfo {author} {\bibfnamefont
  {R.}~\bibnamefont {Trotta}}, \ and\ \bibinfo {author} {\bibfnamefont
  {C.}~\bibnamefont {Weniger}},\ }\href {\doibase
  10.1088/1475-7516/2016/04/037} {\bibfield  {journal} {\bibinfo  {journal}
  {JCAP}\ }\textbf {\bibinfo {volume} {04}},\ \bibinfo {pages} {037} (\bibinfo
  {year} {2016})},\ \Eprint {http://arxiv.org/abs/1507.07008} {arXiv:1507.07008
  [hep-ph]} \BibitemShut {NoStop}%
\bibitem [{\citenamefont {Fonseca}\ \emph {et~al.}(2016)\citenamefont
  {Fonseca}, \citenamefont {Necib},\ and\ \citenamefont
  {Thaler}}]{Fonseca:2015rwa}%
  \BibitemOpen
  \bibfield  {author} {\bibinfo {author} {\bibfnamefont {N.}~\bibnamefont
  {Fonseca}}, \bibinfo {author} {\bibfnamefont {L.}~\bibnamefont {Necib}}, \
  and\ \bibinfo {author} {\bibfnamefont {J.}~\bibnamefont {Thaler}},\ }\href
  {\doibase 10.1088/1475-7516/2016/02/052} {\bibfield  {journal} {\bibinfo
  {journal} {JCAP}\ }\textbf {\bibinfo {volume} {02}},\ \bibinfo {pages} {052}
  (\bibinfo {year} {2016})},\ \Eprint {http://arxiv.org/abs/1507.08295}
  {arXiv:1507.08295 [hep-ph]} \BibitemShut {NoStop}%
\bibitem [{\citenamefont {Freese}\ \emph {et~al.}(2016)\citenamefont {Freese},
  \citenamefont {Lopez}, \citenamefont {Shah},\ and\ \citenamefont
  {Shakya}}]{Freese:2015ysa}%
  \BibitemOpen
  \bibfield  {author} {\bibinfo {author} {\bibfnamefont {K.}~\bibnamefont
  {Freese}}, \bibinfo {author} {\bibfnamefont {A.}~\bibnamefont {Lopez}},
  \bibinfo {author} {\bibfnamefont {N.~R.}\ \bibnamefont {Shah}}, \ and\
  \bibinfo {author} {\bibfnamefont {B.}~\bibnamefont {Shakya}},\ }\href
  {\doibase 10.1007/JHEP04(2016)059} {\bibfield  {journal} {\bibinfo  {journal}
  {JHEP}\ }\textbf {\bibinfo {volume} {04}},\ \bibinfo {pages} {059} (\bibinfo
  {year} {2016})},\ \Eprint {http://arxiv.org/abs/1509.05076} {arXiv:1509.05076
  [hep-ph]} \BibitemShut {NoStop}%
\bibitem [{\citenamefont {Alves}\ \emph {et~al.}(2014)\citenamefont {Alves},
  \citenamefont {Profumo}, \citenamefont {Queiroz},\ and\ \citenamefont
  {Shepherd}}]{Alves:2014yha}%
  \BibitemOpen
  \bibfield  {author} {\bibinfo {author} {\bibfnamefont {A.}~\bibnamefont
  {Alves}}, \bibinfo {author} {\bibfnamefont {S.}~\bibnamefont {Profumo}},
  \bibinfo {author} {\bibfnamefont {F.~S.}\ \bibnamefont {Queiroz}}, \ and\
  \bibinfo {author} {\bibfnamefont {W.}~\bibnamefont {Shepherd}},\ }\href
  {\doibase 10.1103/PhysRevD.90.115003} {\bibfield  {journal} {\bibinfo
  {journal} {Phys. Rev. D}\ }\textbf {\bibinfo {volume} {90}},\ \bibinfo
  {pages} {115003} (\bibinfo {year} {2014})},\ \Eprint
  {http://arxiv.org/abs/1403.5027} {arXiv:1403.5027 [hep-ph]} \BibitemShut
  {NoStop}%
\bibitem [{\citenamefont {Agrawal}\ \emph {et~al.}(2014)\citenamefont
  {Agrawal}, \citenamefont {Batell}, \citenamefont {Hooper},\ and\
  \citenamefont {Lin}}]{Agrawal:2014una}%
  \BibitemOpen
  \bibfield  {author} {\bibinfo {author} {\bibfnamefont {P.}~\bibnamefont
  {Agrawal}}, \bibinfo {author} {\bibfnamefont {B.}~\bibnamefont {Batell}},
  \bibinfo {author} {\bibfnamefont {D.}~\bibnamefont {Hooper}}, \ and\ \bibinfo
  {author} {\bibfnamefont {T.}~\bibnamefont {Lin}},\ }\href {\doibase
  10.1103/PhysRevD.90.063512} {\bibfield  {journal} {\bibinfo  {journal} {Phys.
  Rev. D}\ }\textbf {\bibinfo {volume} {90}},\ \bibinfo {pages} {063512}
  (\bibinfo {year} {2014})},\ \Eprint {http://arxiv.org/abs/1404.1373}
  {arXiv:1404.1373 [hep-ph]} \BibitemShut {NoStop}%
\bibitem [{\citenamefont {Izaguirre}\ \emph {et~al.}(2014)\citenamefont
  {Izaguirre}, \citenamefont {Krnjaic},\ and\ \citenamefont
  {Shuve}}]{Izaguirre:2014vva}%
  \BibitemOpen
  \bibfield  {author} {\bibinfo {author} {\bibfnamefont {E.}~\bibnamefont
  {Izaguirre}}, \bibinfo {author} {\bibfnamefont {G.}~\bibnamefont {Krnjaic}},
  \ and\ \bibinfo {author} {\bibfnamefont {B.}~\bibnamefont {Shuve}},\ }\href
  {\doibase 10.1103/PhysRevD.90.055002} {\bibfield  {journal} {\bibinfo
  {journal} {Phys. Rev. D}\ }\textbf {\bibinfo {volume} {90}},\ \bibinfo
  {pages} {055002} (\bibinfo {year} {2014})},\ \Eprint
  {http://arxiv.org/abs/1404.2018} {arXiv:1404.2018 [hep-ph]} \BibitemShut
  {NoStop}%
\bibitem [{\citenamefont {Ipek}\ \emph {et~al.}(2014)\citenamefont {Ipek},
  \citenamefont {McKeen},\ and\ \citenamefont {Nelson}}]{Ipek:2014gua}%
  \BibitemOpen
  \bibfield  {author} {\bibinfo {author} {\bibfnamefont {S.}~\bibnamefont
  {Ipek}}, \bibinfo {author} {\bibfnamefont {D.}~\bibnamefont {McKeen}}, \ and\
  \bibinfo {author} {\bibfnamefont {A.~E.}\ \bibnamefont {Nelson}},\ }\href
  {\doibase 10.1103/PhysRevD.90.055021} {\bibfield  {journal} {\bibinfo
  {journal} {Phys. Rev. D}\ }\textbf {\bibinfo {volume} {90}},\ \bibinfo
  {pages} {055021} (\bibinfo {year} {2014})},\ \Eprint
  {http://arxiv.org/abs/1404.3716} {arXiv:1404.3716 [hep-ph]} \BibitemShut
  {NoStop}%
\bibitem [{\citenamefont {Tang}\ and\ \citenamefont
  {Zhu}(2016)}]{Tang:2015coo}%
  \BibitemOpen
  \bibfield  {author} {\bibinfo {author} {\bibfnamefont {Y.-L.}\ \bibnamefont
  {Tang}}\ and\ \bibinfo {author} {\bibfnamefont {S.-h.}\ \bibnamefont {Zhu}},\
  }\href {\doibase 10.1007/JHEP03(2016)043} {\bibfield  {journal} {\bibinfo
  {journal} {JHEP}\ }\textbf {\bibinfo {volume} {03}},\ \bibinfo {pages} {043}
  (\bibinfo {year} {2016})},\ \Eprint {http://arxiv.org/abs/1512.02899}
  {arXiv:1512.02899 [hep-ph]} \BibitemShut {NoStop}%
\bibitem [{\citenamefont {Escudero}\ \emph
  {et~al.}(2017{\natexlab{a}})\citenamefont {Escudero}, \citenamefont
  {Hooper},\ and\ \citenamefont {Witte}}]{Escudero:2016kpw}%
  \BibitemOpen
  \bibfield  {author} {\bibinfo {author} {\bibfnamefont {M.}~\bibnamefont
  {Escudero}}, \bibinfo {author} {\bibfnamefont {D.}~\bibnamefont {Hooper}}, \
  and\ \bibinfo {author} {\bibfnamefont {S.~J.}\ \bibnamefont {Witte}},\ }\href
  {\doibase 10.1088/1475-7516/2017/02/038} {\bibfield  {journal} {\bibinfo
  {journal} {JCAP}\ }\textbf {\bibinfo {volume} {02}},\ \bibinfo {pages} {038}
  (\bibinfo {year} {2017}{\natexlab{a}})},\ \Eprint
  {http://arxiv.org/abs/1612.06462} {arXiv:1612.06462 [hep-ph]} \BibitemShut
  {NoStop}%
\bibitem [{\citenamefont {Escudero}\ \emph
  {et~al.}(2017{\natexlab{b}})\citenamefont {Escudero}, \citenamefont {Witte},\
  and\ \citenamefont {Hooper}}]{Escudero:2017yia}%
  \BibitemOpen
  \bibfield  {author} {\bibinfo {author} {\bibfnamefont {M.}~\bibnamefont
  {Escudero}}, \bibinfo {author} {\bibfnamefont {S.~J.}\ \bibnamefont {Witte}},
  \ and\ \bibinfo {author} {\bibfnamefont {D.}~\bibnamefont {Hooper}},\ }\href
  {\doibase 10.1088/1475-7516/2017/11/042} {\bibfield  {journal} {\bibinfo
  {journal} {JCAP}\ }\textbf {\bibinfo {volume} {11}},\ \bibinfo {pages} {042}
  (\bibinfo {year} {2017}{\natexlab{b}})},\ \Eprint
  {http://arxiv.org/abs/1709.07002} {arXiv:1709.07002 [hep-ph]} \BibitemShut
  {NoStop}%
\bibitem [{\citenamefont {Leane}\ and\ \citenamefont
  {Slatyer}(2019)}]{Leane:2019xiy}%
  \BibitemOpen
  \bibfield  {author} {\bibinfo {author} {\bibfnamefont {R.~K.}\ \bibnamefont
  {Leane}}\ and\ \bibinfo {author} {\bibfnamefont {T.~R.}\ \bibnamefont
  {Slatyer}},\ }\href {\doibase 10.1103/PhysRevLett.123.241101} {\bibfield
  {journal} {\bibinfo  {journal} {Phys. Rev. Lett.}\ }\textbf {\bibinfo
  {volume} {123}},\ \bibinfo {pages} {241101} (\bibinfo {year} {2019})},\
  \Eprint {http://arxiv.org/abs/1904.08430} {arXiv:1904.08430 [astro-ph.HE]}
  \BibitemShut {NoStop}%
\bibitem [{\citenamefont {Leane}\ and\ \citenamefont
  {Slatyer}(2020{\natexlab{a}})}]{Leane:2020pfc}%
  \BibitemOpen
  \bibfield  {author} {\bibinfo {author} {\bibfnamefont {R.~K.}\ \bibnamefont
  {Leane}}\ and\ \bibinfo {author} {\bibfnamefont {T.~R.}\ \bibnamefont
  {Slatyer}},\ }\href {\doibase 10.1103/PhysRevD.102.063019} {\bibfield
  {journal} {\bibinfo  {journal} {Phys. Rev. D}\ }\textbf {\bibinfo {volume}
  {102}},\ \bibinfo {pages} {063019} (\bibinfo {year} {2020}{\natexlab{a}})},\
  \Eprint {http://arxiv.org/abs/2002.12371} {arXiv:2002.12371 [astro-ph.HE]}
  \BibitemShut {NoStop}%
\bibitem [{\citenamefont {Leane}\ and\ \citenamefont
  {Slatyer}(2020{\natexlab{b}})}]{Leane:2020nmi}%
  \BibitemOpen
  \bibfield  {author} {\bibinfo {author} {\bibfnamefont {R.~K.}\ \bibnamefont
  {Leane}}\ and\ \bibinfo {author} {\bibfnamefont {T.~R.}\ \bibnamefont
  {Slatyer}},\ }\href {\doibase 10.1103/PhysRevLett.125.121105} {\bibfield
  {journal} {\bibinfo  {journal} {Phys. Rev. Lett.}\ }\textbf {\bibinfo
  {volume} {125}},\ \bibinfo {pages} {121105} (\bibinfo {year}
  {2020}{\natexlab{b}})},\ \Eprint {http://arxiv.org/abs/2002.12370}
  {arXiv:2002.12370 [astro-ph.HE]} \BibitemShut {NoStop}%
\bibitem [{\citenamefont {Zhong}\ \emph {et~al.}(2020)\citenamefont {Zhong},
  \citenamefont {McDermott}, \citenamefont {Cholis},\ and\ \citenamefont
  {Fox}}]{Zhong:2019ycb}%
  \BibitemOpen
  \bibfield  {author} {\bibinfo {author} {\bibfnamefont {Y.-M.}\ \bibnamefont
  {Zhong}}, \bibinfo {author} {\bibfnamefont {S.~D.}\ \bibnamefont
  {McDermott}}, \bibinfo {author} {\bibfnamefont {I.}~\bibnamefont {Cholis}}, \
  and\ \bibinfo {author} {\bibfnamefont {P.~J.}\ \bibnamefont {Fox}},\ }\href
  {\doibase 10.1103/PhysRevLett.124.231103} {\bibfield  {journal} {\bibinfo
  {journal} {Phys. Rev. Lett.}\ }\textbf {\bibinfo {volume} {124}},\ \bibinfo
  {pages} {231103} (\bibinfo {year} {2020})},\ \Eprint
  {http://arxiv.org/abs/1911.12369} {arXiv:1911.12369 [astro-ph.HE]}
  \BibitemShut {NoStop}%
\bibitem [{\citenamefont {List}\ \emph {et~al.}(2020)\citenamefont {List},
  \citenamefont {Rodd}, \citenamefont {Lewis},\ and\ \citenamefont
  {Bhat}}]{List:2020mzd}%
  \BibitemOpen
  \bibfield  {author} {\bibinfo {author} {\bibfnamefont {F.}~\bibnamefont
  {List}}, \bibinfo {author} {\bibfnamefont {N.~L.}\ \bibnamefont {Rodd}},
  \bibinfo {author} {\bibfnamefont {G.~F.}\ \bibnamefont {Lewis}}, \ and\
  \bibinfo {author} {\bibfnamefont {I.}~\bibnamefont {Bhat}},\ }\href {\doibase
  10.1103/PhysRevLett.125.241102} {\bibfield  {journal} {\bibinfo  {journal}
  {Phys. Rev. Lett.}\ }\textbf {\bibinfo {volume} {125}},\ \bibinfo {pages}
  {241102} (\bibinfo {year} {2020})},\ \Eprint
  {http://arxiv.org/abs/2006.12504} {arXiv:2006.12504 [astro-ph.HE]}
  \BibitemShut {NoStop}%
\bibitem [{\citenamefont {Calore}\ \emph {et~al.}(2021)\citenamefont {Calore},
  \citenamefont {Donato},\ and\ \citenamefont {Manconi}}]{Calore:2021bty}%
  \BibitemOpen
  \bibfield  {author} {\bibinfo {author} {\bibfnamefont {F.}~\bibnamefont
  {Calore}}, \bibinfo {author} {\bibfnamefont {F.}~\bibnamefont {Donato}}, \
  and\ \bibinfo {author} {\bibfnamefont {S.}~\bibnamefont {Manconi}},\
  }\href@noop {} {\  (\bibinfo {year} {2021})},\ \Eprint
  {http://arxiv.org/abs/2102.12497} {arXiv:2102.12497 [astro-ph.HE]}
  \BibitemShut {NoStop}%
\bibitem [{\citenamefont {Macias}\ \emph {et~al.}(2018)\citenamefont {Macias},
  \citenamefont {Gordon}, \citenamefont {Crocker}, \citenamefont {Coleman},
  \citenamefont {Paterson}, \citenamefont {Horiuchi},\ and\ \citenamefont
  {Pohl}}]{Macias:2016nev}%
  \BibitemOpen
  \bibfield  {author} {\bibinfo {author} {\bibfnamefont {O.}~\bibnamefont
  {Macias}}, \bibinfo {author} {\bibfnamefont {C.}~\bibnamefont {Gordon}},
  \bibinfo {author} {\bibfnamefont {R.~M.}\ \bibnamefont {Crocker}}, \bibinfo
  {author} {\bibfnamefont {B.}~\bibnamefont {Coleman}}, \bibinfo {author}
  {\bibfnamefont {D.}~\bibnamefont {Paterson}}, \bibinfo {author}
  {\bibfnamefont {S.}~\bibnamefont {Horiuchi}}, \ and\ \bibinfo {author}
  {\bibfnamefont {M.}~\bibnamefont {Pohl}},\ }\href {\doibase
  10.1038/s41550-018-0414-3} {\bibfield  {journal} {\bibinfo  {journal} {Nature
  Astron.}\ }\textbf {\bibinfo {volume} {2}},\ \bibinfo {pages} {387} (\bibinfo
  {year} {2018})},\ \Eprint {http://arxiv.org/abs/1611.06644} {arXiv:1611.06644
  [astro-ph.HE]} \BibitemShut {NoStop}%
\bibitem [{\citenamefont {Bartels}\ \emph
  {et~al.}(2018{\natexlab{a}})\citenamefont {Bartels}, \citenamefont {Storm},
  \citenamefont {Weniger},\ and\ \citenamefont {Calore}}]{Bartels:2017vsx}%
  \BibitemOpen
  \bibfield  {author} {\bibinfo {author} {\bibfnamefont {R.}~\bibnamefont
  {Bartels}}, \bibinfo {author} {\bibfnamefont {E.}~\bibnamefont {Storm}},
  \bibinfo {author} {\bibfnamefont {C.}~\bibnamefont {Weniger}}, \ and\
  \bibinfo {author} {\bibfnamefont {F.}~\bibnamefont {Calore}},\ }\href
  {\doibase 10.1038/s41550-018-0531-z} {\bibfield  {journal} {\bibinfo
  {journal} {Nature Astron.}\ }\textbf {\bibinfo {volume} {2}},\ \bibinfo
  {pages} {819} (\bibinfo {year} {2018}{\natexlab{a}})},\ \Eprint
  {http://arxiv.org/abs/1711.04778} {arXiv:1711.04778 [astro-ph.HE]}
  \BibitemShut {NoStop}%
\bibitem [{\citenamefont {Macias}\ \emph {et~al.}(2019)\citenamefont {Macias},
  \citenamefont {Horiuchi}, \citenamefont {Kaplinghat}, \citenamefont {Gordon},
  \citenamefont {Crocker},\ and\ \citenamefont {Nataf}}]{Macias:2019omb}%
  \BibitemOpen
  \bibfield  {author} {\bibinfo {author} {\bibfnamefont {O.}~\bibnamefont
  {Macias}}, \bibinfo {author} {\bibfnamefont {S.}~\bibnamefont {Horiuchi}},
  \bibinfo {author} {\bibfnamefont {M.}~\bibnamefont {Kaplinghat}}, \bibinfo
  {author} {\bibfnamefont {C.}~\bibnamefont {Gordon}}, \bibinfo {author}
  {\bibfnamefont {R.~M.}\ \bibnamefont {Crocker}}, \ and\ \bibinfo {author}
  {\bibfnamefont {D.~M.}\ \bibnamefont {Nataf}},\ }\href {\doibase
  10.1088/1475-7516/2019/09/042} {\bibfield  {journal} {\bibinfo  {journal}
  {JCAP}\ }\textbf {\bibinfo {volume} {09}},\ \bibinfo {pages} {042} (\bibinfo
  {year} {2019})},\ \Eprint {http://arxiv.org/abs/1901.03822} {arXiv:1901.03822
  [astro-ph.HE]} \BibitemShut {NoStop}%
\bibitem [{\citenamefont {Cholis}\ \emph {et~al.}(2015)\citenamefont {Cholis},
  \citenamefont {Hooper},\ and\ \citenamefont {Linden}}]{Cholis:2014lta}%
  \BibitemOpen
  \bibfield  {author} {\bibinfo {author} {\bibfnamefont {I.}~\bibnamefont
  {Cholis}}, \bibinfo {author} {\bibfnamefont {D.}~\bibnamefont {Hooper}}, \
  and\ \bibinfo {author} {\bibfnamefont {T.}~\bibnamefont {Linden}},\ }\href
  {\doibase 10.1088/1475-7516/2015/06/043} {\bibfield  {journal} {\bibinfo
  {journal} {JCAP}\ }\textbf {\bibinfo {volume} {06}},\ \bibinfo {pages} {043}
  (\bibinfo {year} {2015})},\ \Eprint {http://arxiv.org/abs/1407.5625}
  {arXiv:1407.5625 [astro-ph.HE]} \BibitemShut {NoStop}%
\bibitem [{\citenamefont {Hooper}\ and\ \citenamefont
  {Mohlabeng}(2016)}]{Hooper:2015jlu}%
  \BibitemOpen
  \bibfield  {author} {\bibinfo {author} {\bibfnamefont {D.}~\bibnamefont
  {Hooper}}\ and\ \bibinfo {author} {\bibfnamefont {G.}~\bibnamefont
  {Mohlabeng}},\ }\href {\doibase 10.1088/1475-7516/2016/03/049} {\bibfield
  {journal} {\bibinfo  {journal} {JCAP}\ }\textbf {\bibinfo {volume} {03}},\
  \bibinfo {pages} {049} (\bibinfo {year} {2016})},\ \Eprint
  {http://arxiv.org/abs/1512.04966} {arXiv:1512.04966 [astro-ph.HE]}
  \BibitemShut {NoStop}%
\bibitem [{\citenamefont {Bartels}\ \emph
  {et~al.}(2018{\natexlab{b}})\citenamefont {Bartels}, \citenamefont {Hooper},
  \citenamefont {Linden}, \citenamefont {Mishra-Sharma}, \citenamefont {Rodd},
  \citenamefont {Safdi},\ and\ \citenamefont {Slatyer}}]{Bartels:2017xba}%
  \BibitemOpen
  \bibfield  {author} {\bibinfo {author} {\bibfnamefont {R.}~\bibnamefont
  {Bartels}}, \bibinfo {author} {\bibfnamefont {D.}~\bibnamefont {Hooper}},
  \bibinfo {author} {\bibfnamefont {T.}~\bibnamefont {Linden}}, \bibinfo
  {author} {\bibfnamefont {S.}~\bibnamefont {Mishra-Sharma}}, \bibinfo {author}
  {\bibfnamefont {N.~L.}\ \bibnamefont {Rodd}}, \bibinfo {author}
  {\bibfnamefont {B.~R.}\ \bibnamefont {Safdi}}, \ and\ \bibinfo {author}
  {\bibfnamefont {T.~R.}\ \bibnamefont {Slatyer}},\ }\href {\doibase
  10.1016/j.dark.2018.04.004} {\bibfield  {journal} {\bibinfo  {journal} {Phys.
  Dark Univ.}\ }\textbf {\bibinfo {volume} {20}},\ \bibinfo {pages} {88}
  (\bibinfo {year} {2018}{\natexlab{b}})},\ \Eprint
  {http://arxiv.org/abs/1710.10266} {arXiv:1710.10266 [astro-ph.HE]}
  \BibitemShut {NoStop}%
\bibitem [{\citenamefont {Hooper}\ and\ \citenamefont
  {Linden}(2016)}]{Hooper:2016rap}%
  \BibitemOpen
  \bibfield  {author} {\bibinfo {author} {\bibfnamefont {D.}~\bibnamefont
  {Hooper}}\ and\ \bibinfo {author} {\bibfnamefont {T.}~\bibnamefont
  {Linden}},\ }\href {\doibase 10.1088/1475-7516/2016/08/018} {\bibfield
  {journal} {\bibinfo  {journal} {JCAP}\ }\textbf {\bibinfo {volume} {08}},\
  \bibinfo {pages} {018} (\bibinfo {year} {2016})},\ \Eprint
  {http://arxiv.org/abs/1606.09250} {arXiv:1606.09250 [astro-ph.HE]}
  \BibitemShut {NoStop}%
\bibitem [{\citenamefont {Haggard}\ \emph {et~al.}(2017)\citenamefont
  {Haggard}, \citenamefont {Heinke}, \citenamefont {Hooper},\ and\
  \citenamefont {Linden}}]{Haggard:2017lyq}%
  \BibitemOpen
  \bibfield  {author} {\bibinfo {author} {\bibfnamefont {D.}~\bibnamefont
  {Haggard}}, \bibinfo {author} {\bibfnamefont {C.}~\bibnamefont {Heinke}},
  \bibinfo {author} {\bibfnamefont {D.}~\bibnamefont {Hooper}}, \ and\ \bibinfo
  {author} {\bibfnamefont {T.}~\bibnamefont {Linden}},\ }\href {\doibase
  10.1088/1475-7516/2017/05/056} {\bibfield  {journal} {\bibinfo  {journal}
  {JCAP}\ }\textbf {\bibinfo {volume} {05}},\ \bibinfo {pages} {056} (\bibinfo
  {year} {2017})},\ \Eprint {http://arxiv.org/abs/1701.02726} {arXiv:1701.02726
  [astro-ph.HE]} \BibitemShut {NoStop}%
\bibitem [{\citenamefont {Calore}\ \emph {et~al.}(2016)\citenamefont {Calore},
  \citenamefont {Di~Mauro}, \citenamefont {Donato}, \citenamefont {Hessels},\
  and\ \citenamefont {Weniger}}]{Calore:2015bsx}%
  \BibitemOpen
  \bibfield  {author} {\bibinfo {author} {\bibfnamefont {F.}~\bibnamefont
  {Calore}}, \bibinfo {author} {\bibfnamefont {M.}~\bibnamefont {Di~Mauro}},
  \bibinfo {author} {\bibfnamefont {F.}~\bibnamefont {Donato}}, \bibinfo
  {author} {\bibfnamefont {J.~W.~T.}\ \bibnamefont {Hessels}}, \ and\ \bibinfo
  {author} {\bibfnamefont {C.}~\bibnamefont {Weniger}},\ }\href {\doibase
  10.3847/0004-637X/827/2/143} {\bibfield  {journal} {\bibinfo  {journal}
  {Astrophys. J.}\ }\textbf {\bibinfo {volume} {827}},\ \bibinfo {pages} {143}
  (\bibinfo {year} {2016})},\ \Eprint {http://arxiv.org/abs/1512.06825}
  {arXiv:1512.06825 [astro-ph.HE]} \BibitemShut {NoStop}%
\bibitem [{\citenamefont {Albert}\ \emph {et~al.}(2017)\citenamefont {Albert}
  \emph {et~al.}}]{Fermi-LAT:2016uux}%
  \BibitemOpen
  \bibfield  {author} {\bibinfo {author} {\bibfnamefont {A.}~\bibnamefont
  {Albert}} \emph {et~al.} (\bibinfo {collaboration} {Fermi-LAT, DES}),\ }\href
  {\doibase 10.3847/1538-4357/834/2/110} {\bibfield  {journal} {\bibinfo
  {journal} {Astrophys. J.}\ }\textbf {\bibinfo {volume} {834}},\ \bibinfo
  {pages} {110} (\bibinfo {year} {2017})},\ \Eprint
  {http://arxiv.org/abs/1611.03184} {arXiv:1611.03184 [astro-ph.HE]}
  \BibitemShut {NoStop}%
\bibitem [{\citenamefont {Cuoco}\ \emph {et~al.}(2017)\citenamefont {Cuoco},
  \citenamefont {Kr\"amer},\ and\ \citenamefont {Korsmeier}}]{Cuoco:2016eej}%
  \BibitemOpen
  \bibfield  {author} {\bibinfo {author} {\bibfnamefont {A.}~\bibnamefont
  {Cuoco}}, \bibinfo {author} {\bibfnamefont {M.}~\bibnamefont {Kr\"amer}}, \
  and\ \bibinfo {author} {\bibfnamefont {M.}~\bibnamefont {Korsmeier}},\ }\href
  {\doibase 10.1103/PhysRevLett.118.191102} {\bibfield  {journal} {\bibinfo
  {journal} {Phys. Rev. Lett.}\ }\textbf {\bibinfo {volume} {118}},\ \bibinfo
  {pages} {191102} (\bibinfo {year} {2017})},\ \Eprint
  {http://arxiv.org/abs/1610.03071} {arXiv:1610.03071 [astro-ph.HE]}
  \BibitemShut {NoStop}%
\bibitem [{\citenamefont {Cui}\ \emph {et~al.}(2017)\citenamefont {Cui},
  \citenamefont {Yuan}, \citenamefont {Tsai},\ and\ \citenamefont
  {Fan}}]{Cui:2016ppb}%
  \BibitemOpen
  \bibfield  {author} {\bibinfo {author} {\bibfnamefont {M.-Y.}\ \bibnamefont
  {Cui}}, \bibinfo {author} {\bibfnamefont {Q.}~\bibnamefont {Yuan}}, \bibinfo
  {author} {\bibfnamefont {Y.-L.~S.}\ \bibnamefont {Tsai}}, \ and\ \bibinfo
  {author} {\bibfnamefont {Y.-Z.}\ \bibnamefont {Fan}},\ }\href {\doibase
  10.1103/PhysRevLett.118.191101} {\bibfield  {journal} {\bibinfo  {journal}
  {Phys. Rev. Lett.}\ }\textbf {\bibinfo {volume} {118}},\ \bibinfo {pages}
  {191101} (\bibinfo {year} {2017})},\ \Eprint
  {http://arxiv.org/abs/1610.03840} {arXiv:1610.03840 [astro-ph.HE]}
  \BibitemShut {NoStop}%
\bibitem [{\citenamefont {Cholis}\ \emph {et~al.}(2019)\citenamefont {Cholis},
  \citenamefont {Linden},\ and\ \citenamefont {Hooper}}]{Cholis:2019ejx}%
  \BibitemOpen
  \bibfield  {author} {\bibinfo {author} {\bibfnamefont {I.}~\bibnamefont
  {Cholis}}, \bibinfo {author} {\bibfnamefont {T.}~\bibnamefont {Linden}}, \
  and\ \bibinfo {author} {\bibfnamefont {D.}~\bibnamefont {Hooper}},\ }\href
  {\doibase 10.1103/PhysRevD.99.103026} {\bibfield  {journal} {\bibinfo
  {journal} {Phys. Rev. D}\ }\textbf {\bibinfo {volume} {99}},\ \bibinfo
  {pages} {103026} (\bibinfo {year} {2019})},\ \Eprint
  {http://arxiv.org/abs/1903.02549} {arXiv:1903.02549 [astro-ph.HE]}
  \BibitemShut {NoStop}%
\bibitem [{\citenamefont {Cuoco}\ \emph {et~al.}(2019)\citenamefont {Cuoco},
  \citenamefont {Heisig}, \citenamefont {Klamt}, \citenamefont {Korsmeier},\
  and\ \citenamefont {Kr\"amer}}]{Cuoco:2019kuu}%
  \BibitemOpen
  \bibfield  {author} {\bibinfo {author} {\bibfnamefont {A.}~\bibnamefont
  {Cuoco}}, \bibinfo {author} {\bibfnamefont {J.}~\bibnamefont {Heisig}},
  \bibinfo {author} {\bibfnamefont {L.}~\bibnamefont {Klamt}}, \bibinfo
  {author} {\bibfnamefont {M.}~\bibnamefont {Korsmeier}}, \ and\ \bibinfo
  {author} {\bibfnamefont {M.}~\bibnamefont {Kr\"amer}},\ }\href {\doibase
  10.1103/PhysRevD.99.103014} {\bibfield  {journal} {\bibinfo  {journal} {Phys.
  Rev. D}\ }\textbf {\bibinfo {volume} {99}},\ \bibinfo {pages} {103014}
  (\bibinfo {year} {2019})},\ \Eprint {http://arxiv.org/abs/1903.01472}
  {arXiv:1903.01472 [astro-ph.HE]} \BibitemShut {NoStop}%
\bibitem [{\citenamefont {Abramowski}\ \emph {et~al.}(2016)\citenamefont
  {Abramowski} \emph {et~al.}}]{Abramowski:2016mir}%
  \BibitemOpen
  \bibfield  {author} {\bibinfo {author} {\bibfnamefont {A.}~\bibnamefont
  {Abramowski}} \emph {et~al.} (\bibinfo {collaboration} {H.E.S.S.}),\ }\href
  {\doibase 10.1038/nature17147} {\bibfield  {journal} {\bibinfo  {journal}
  {Nature}\ }\textbf {\bibinfo {volume} {531}},\ \bibinfo {pages} {476}
  (\bibinfo {year} {2016})},\ \Eprint {http://arxiv.org/abs/1603.07730}
  {arXiv:1603.07730 [astro-ph.HE]} \BibitemShut {NoStop}%
\bibitem [{\citenamefont {Hooper}\ \emph {et~al.}(2018)\citenamefont {Hooper},
  \citenamefont {Cholis},\ and\ \citenamefont {Linden}}]{Hooper:2017rzt}%
  \BibitemOpen
  \bibfield  {author} {\bibinfo {author} {\bibfnamefont {D.}~\bibnamefont
  {Hooper}}, \bibinfo {author} {\bibfnamefont {I.}~\bibnamefont {Cholis}}, \
  and\ \bibinfo {author} {\bibfnamefont {T.}~\bibnamefont {Linden}},\ }\href
  {\doibase 10.1016/j.dark.2018.05.004} {\bibfield  {journal} {\bibinfo
  {journal} {Phys. Dark Univ.}\ }\textbf {\bibinfo {volume} {21}},\ \bibinfo
  {pages} {40} (\bibinfo {year} {2018})},\ \Eprint
  {http://arxiv.org/abs/1705.09293} {arXiv:1705.09293 [astro-ph.HE]}
  \BibitemShut {NoStop}%
\bibitem [{\citenamefont {Song}\ \emph {et~al.}(2019)\citenamefont {Song},
  \citenamefont {Macias},\ and\ \citenamefont {Horiuchi}}]{Song:2019nrx}%
  \BibitemOpen
  \bibfield  {author} {\bibinfo {author} {\bibfnamefont {D.}~\bibnamefont
  {Song}}, \bibinfo {author} {\bibfnamefont {O.}~\bibnamefont {Macias}}, \ and\
  \bibinfo {author} {\bibfnamefont {S.}~\bibnamefont {Horiuchi}},\ }\href
  {\doibase 10.1103/PhysRevD.99.123020} {\bibfield  {journal} {\bibinfo
  {journal} {Phys. Rev. D}\ }\textbf {\bibinfo {volume} {99}},\ \bibinfo
  {pages} {123020} (\bibinfo {year} {2019})},\ \Eprint
  {http://arxiv.org/abs/1901.07025} {arXiv:1901.07025 [astro-ph.HE]}
  \BibitemShut {NoStop}%
\bibitem [{\citenamefont {Acharya}\ \emph {et~al.}(2018)\citenamefont {Acharya}
  \emph {et~al.}}]{CTAConsortium:2018tzg}%
  \BibitemOpen
  \bibfield  {author} {\bibinfo {author} {\bibfnamefont {B.}~\bibnamefont
  {Acharya}} \emph {et~al.} (\bibinfo {collaboration} {CTA Consortium}),\
  }\href {\doibase 10.1142/10986} {\emph {\bibinfo {title} {{Science with the
  Cherenkov Telescope Array}}}}\ (\bibinfo  {publisher} {WSP},\ \bibinfo {year}
  {2018})\ \Eprint {http://arxiv.org/abs/1709.07997} {arXiv:1709.07997
  [astro-ph.IM]} \BibitemShut {NoStop}%
\bibitem [{\citenamefont {Macias}\ \emph {et~al.}(2021)\citenamefont {Macias},
  \citenamefont {van Leijen}, \citenamefont {Song}, \citenamefont {Ando},
  \citenamefont {Horiuchi},\ and\ \citenamefont {Crocker}}]{Macias:2021boz}%
  \BibitemOpen
  \bibfield  {author} {\bibinfo {author} {\bibfnamefont {O.}~\bibnamefont
  {Macias}}, \bibinfo {author} {\bibfnamefont {H.}~\bibnamefont {van Leijen}},
  \bibinfo {author} {\bibfnamefont {D.}~\bibnamefont {Song}}, \bibinfo {author}
  {\bibfnamefont {S.}~\bibnamefont {Ando}}, \bibinfo {author} {\bibfnamefont
  {S.}~\bibnamefont {Horiuchi}}, \ and\ \bibinfo {author} {\bibfnamefont
  {R.~M.}\ \bibnamefont {Crocker}},\ }\href@noop {} {\  (\bibinfo {year}
  {2021})},\ \Eprint {http://arxiv.org/abs/2102.05648} {arXiv:2102.05648
  [astro-ph.HE]} \BibitemShut {NoStop}%
\bibitem [{\citenamefont {Sudoh}\ \emph {et~al.}(2020)\citenamefont {Sudoh},
  \citenamefont {Linden},\ and\ \citenamefont {Beacom}}]{Sudoh:2020hyu}%
  \BibitemOpen
  \bibfield  {author} {\bibinfo {author} {\bibfnamefont {T.}~\bibnamefont
  {Sudoh}}, \bibinfo {author} {\bibfnamefont {T.}~\bibnamefont {Linden}}, \
  and\ \bibinfo {author} {\bibfnamefont {J.~F.}\ \bibnamefont {Beacom}},\
  }\href@noop {} {\  (\bibinfo {year} {2020})},\ \Eprint
  {http://arxiv.org/abs/2005.08982} {arXiv:2005.08982 [astro-ph.GA]}
  \BibitemShut {NoStop}%
\end{thebibliography}%

\end{document}